\documentstyle[11pt]{article}
\newcommand{\bigzerou}{%
\smash{\lower1.7ex\hbox{\bg 0}}}
\renewcommand{\theequation}{\arabic{section}.\arabic{equation}}
\newcommand{\beq}{\begin{equation}}
\newcommand{\enq}{\end{equation}}

\newcommand{\mapright}[1]{%
\smash{\mathop{%
\hbox to 1.0cm{\rightarrowfill}}\limits^{#1}}}
\newcommand{\mapleft}[1]{%
\smash{\mathop{%
\hbox to 1.3cm{\leftarrowfill}}\limits^{#1}}}

\newcommand{\no}{\nonumber}

\newtheorem{thm}{Theorem}
\newtheorem{conj}{Conjecture}

\newcommand{\beqy}{\begin{eqnarray}}
\newcommand{\enqy}{\end{eqnarray}}

\begin{document}

%%%%%%%%%%%%%%%%%%%title%%%%%%%%%%%
%%%%%%%%%%%%%%%%%%%%%%%%%%%%%%%%%%%

\begin{titlepage}
\vglue 3cm

\begin{center}
\vglue 0.5cm
{\Large\bf Affine Lie Algebras and $S$-Duality of ${\cal N}=4$ Super Yang-Mills Theory for $ADE$ Gauge Groups on $K3$}
\vglue 1cm
{\large Toru Sasaki}
\vglue 0.5cm
{\it Graduate School of Mathematics,
Nagoya University, Nagoya, 464-8602, Japan}\\
{x03001e@math.nagoya-u.ac.jp}

\baselineskip=12pt

\vglue 1cm
\begin{abstract}
We attempt to determine the partition function of ${\cal N}=4$ super Yang-Mills theory for $ADE$ gauge groups on $K3$ and investigate the relation with affine Lie algebras.
In particular we describe eta functions, which compose $SU(N)$ partition function, by level $N~A_{N-1}$ theta functions.
Moreover we find $D,E$ theta functions,
which satisfy the Montonen-Olive duality for $D,E$ partition functions.

\end{abstract}
\end{center}
\end{titlepage}

\section{Introduction}
\setcounter{equation}{0}
This article is one of a series of our trials to determine ${\cal N}=4~ADE$
partition function on $K3$ \cite{jin3,sasaki1,sasaki2}.
In this article we attempt to reveal the relation between ${\cal N}=4~ADE$ partition function on $K3$ and affine Lie algebra.
We provide the identity between eta function $1/\eta(\tau/N)$ and level $N~A_{N-1}$ theta functions over $\eta(\tau)^N$,
$\theta_{A_{N-1}}^\beta(\tau)/\eta(\tau)^N$, which is a kind of blow-up formula\cite{yoshioka, kap,jin2}.
We consider that $SU(N)$ partition function on $K3$ is described by this $SU(N)$ blow-up formula and attempt to generalize this to $D,E$ case.

${\cal N}=4$ supersymmetric theory is one of good laboratories of	studying SUSY duality.
In particular on 4-K\"ahler manifold, due to topological twisting
we obtain twisted ${\cal N}=4$ super Yang-Mills theory, which is a kind of topological field theory (so called Vafa-Witten theory)\cite{vafa-witten,laba}.
Of course, twisted ${\cal N}=4$ super Yang-Mills theory is not ${\cal N}=4$ supersymmetric theory on original 4-manifold itself, but is providing interesting topics associated with SUSY duality \cite{bonelli,E-S,iqbal,laba,lozano,m-v,yoshi}.
Typical properties of Vafa-Witten theory are summarized
in the following Vafa-Witten conjectures (more detail explanation is done in the next section)\cite{vafa-witten, lozano, yoshi}.\\
(I) The partition function of twisted ${\cal N}=4$ super Yang-Mills theory is the generating function of the Euler number of the instanton moduli space. \\(II) This partition function satisfies Montonen-Olive duality ($S$-duality of ${\cal N}=4$ SUSY)\cite{M-O}.
(III) Because of A-H-S dimension formula\cite{AHS}, the instanton moduli space of small instanton number must vanish
and hence so the corresponding $q$ terms of the partition function. This condition is called gap condition. \\
Each conjecture is challenging problem.
However we ask how we can determine the partition function of twisted ${\cal N}=4$ super Yang-Mills theory if the partition function satisfies above conjectures.

As for twisted ${\cal N}=4~SU(N)$ theory on some K\"ahler 4-manifolds, partition functions satisfying above conjectures were already derived \cite{vafa-witten, m-v}.
However there has been no trail to determine $ADE$ partition function,
although there was a conjecture for $ADE$ gauge groups supporting
the Montonen-Olive duality conjecture \cite{vafa}.
Here we attempt to determine ${\cal N}=4$ partition function for $ADE$ gauge group ${\cal G}_r$ on $K3$ so that it satisfies the Vafa-Witten conjecture. We had taken two approaches to this aim.
The first approach is the straightforward generalization of the $SU(N)$ case. We found the identity
$1/\eta(\tau/N)=\theta_{A_{N-1}}(\tau)/\eta(\tau)^N$ \cite{jin2}.
Here $1/\eta(\tau/N)$ is the building block of $SU(N)$ partition function on $K3$ and $\theta_{A_{N-1}}(\tau)$ is a kind of $A_{N-1}$ theta function. In \cite{jin3} we generalized $\theta_{A_{N-1}}(\tau)/\eta(\tau)^N$ to $\theta_{{\cal G}_r}(\tau)/\eta(\tau)^{r+1}$ and constructed  the ${\cal G}_r$ partition function on $K3$ so that it satisfies the Montonen-Olive duality.
The second approach is to describe partition function by a kind of Hecke operator. In \cite{sasaki1} we invented the deformed Hecke operator,
which satisfies the Montonen-Olive duality automatically.
We only had to determine the function so that the operated function satisfies the gap condition.
These two approaches have both advantages and disadvantages,
and do not provide complete partition function
for all $ADE$ gauge groups
satisfying all Vafa-Witten conjectures.
The first approach covers all $ADE$ gauge groups, but its partition function does not satisfy the gap condition in general.
The second approach can provide the partition function satisfying all Vafa-Witten conjectures, but cannot cover all $ADE$ gauge groups.

In this article we take the first approach and search for the method of generalization to $D,E$ case.
The starting point is the identity:
\beq
\frac{1}{\eta(\frac{\tau}{N})}=\frac{\theta_{A_{N-1}}(\tau)}{\eta(\tau)^N},
\enq
\beq
\theta_{A_{N-1}}(\tau)=\sum_{\beta\in \frac{M^*}{NM}}n_\beta\theta_{A_{N-1}}^\beta(\tau),{}^\exists n_\beta\in {\bf Z},
\enq
where $\theta_{A_{N-1}}^\beta(\tau)$ is a level $N~A_{N-1}$theta function with weight vector $\beta\in \frac{M^*}{NM}$. $M,M^*$ are root lattice and dual lattice respectively \cite{kac, wakimoto1}. This equality had already appeared in \cite{jin2},where this formula is interpreted as ${\cal O}(-N)$ blow-up formula.
Here ${\cal O}(-N)$ stands for ${\cal O}(-N)$ curve whose intersection number is $-N$.
In the first half of this article, we verify this identity for small rank. This identity means that the partition function is described by blow-up formula using level $N~A_{N-1}$ theta functions.
In the second half of this article, we introduce the analogue for $D,E$ case, by using level $|\Gamma_{\cal G}|~{\cal G}_r$ theta functions.
Here $\Gamma_{\cal G}$ is center of $\Gamma_{\cal G}$ and $|\Gamma_{\cal G}|$ is the number of elements of it.
Taking account of the modular property of the ${\cal G}_r$ partition function, which is the same as that of $SU(|\Gamma_{\cal G}|)$ partition function,
we search for the ${\cal G}_r$ theta function by using level $|\Gamma_{\cal G}|~{\cal G}_r$ theta functions whose modular property is the same as that of $\eta(\tau)^{r+1}/\eta(\tau/|\Gamma_{\cal G}|)$.
Once we determine the $D,E$ theta function, it is straightforward to obtain the $D,E$ partition functions on $K3$ satisfying Montonen-Olive duality.

The organization of this article is the following:
In Sec.2 we review the Vafa-Witten theory and the Vafa-Witten conjecture
for the twisted ${\cal N}=4$ super Yang-Mills theory.
In Sec.3 we review our previous works for determining ${\cal N}=4~ADE$ partition function on $K3$.
In Sec.4 we provide the identity between $1/\eta(\tau/N)$ and
level $N~A_{N-1}$ theta functions over $\eta(\tau)^N$.
We also verify this identity for small rank.
In Sec.5 we introduce ${\cal G}_r$ theta function by using level $|\Gamma_{\cal G}|~{\cal G}_r$ theta functions, whose modular property is the same as $\eta(\tau)^{r+1}/\eta(\tau/|\Gamma_{\cal G}|)$.
In Sec.6. we conclude and discuss the remaining problems.

\section{Review of Vafa-Witten Theory}
\setcounter{equation}{0}
Vafa and Witten have first intended to test $S$-duality conjecture of ${\cal N}=4$ super Yang-Mills
theory, by determining the exact partition functions for the twisted theory \cite{vafa-witten}.
Although ${\cal N}=4$ super Yang-Mills theory on $4$-manifold
is exactly finite and conformally invariant,
it is still hard to determine the exact ${\cal N}=4$ partition function itself.
However the twisted ${\cal N}=4$ super Yang-Mills theory is topological field theory \cite{laba}, and its
partition function can be determined exactly.
In this context, Vafa and Witten
investigated the mathematical results of
the $SU(N)$ partition functions on simple
K\"ahler 4-manifolds such as $K3, C{\bf P}^2, C{\hat {\bf P}^2}$(blow-uped $C{\bf P}^2)$ and $ALE$ \cite{vafa-witten}.
On the other hand,
$S$-duality conjecture of (twisted)${\cal N}=4$ super Yang-Mills theory is believed to be the one considered by Montonen and Olive \cite{M-O}.
This $S$-duality accompanies the exchange of gauge groups ${\cal G} \leftrightarrow{\hat {\cal G}} $.
${\hat {\cal G}}$ is the dual of ${\cal G}$.
If	${\cal G}$ is $ADE$ gauge group, ${\hat {\cal G}}$ is given by ${\hat {\cal G}}={\cal G}/{\Gamma_{\cal G}}$.
Here
$\Gamma_{\cal G}$ is the center of ${\cal G}$  and $|\Gamma_{\cal G}|$ is the number of elements of $\Gamma_{\cal G}$.
To classify the theory for ${\cal G}/\Gamma_{\cal G}$,
we introduce 't Hooft fluxes $v\in H^2(X,\Gamma_{\cal G})$.
We point out that ${\cal G}/\Gamma_{\cal G}$ with $v=0$ is regarded as ${\cal G}$ itself \cite{jin3}.
\begin{table}[h]
\begin{center}
\begin{tabular}{|c|c|c|c|}
\hline
${\cal G}$& $\Gamma_{\cal G}$ & $|\Gamma_{\cal G}|$ \\
\hline
$A_{N-1}$ & ${\bf Z}_N$ &N \\
\hline
$D_{2N}$ &	${\bf Z}_2\times{\bf Z}_2$ &4 \\
\hline
$D_{2N+1}$ &  ${\bf Z}_4$ &4 \\
\hline
$E_r, r=6,7,8$ &  ${\bf Z}_{9-r}$ &$9-r$ \\
%\hline
%$E_7$ &  ${\bf Z}_2$\\
%\hline
%$E_8$ &  trivial\\
\hline
\end{tabular}
\end{center}
\end{table}

\paragraph{Vafa-Witten Conjecture}~\\
In this part we provide the celebrated Vafa-Witten conjecture
concerned with partition function of twisted ${\cal N}=4$ super Yang-Mills theory
on 4-manifold $X$ \cite{vafa-witten, lozano, yoshi}.
Vafa and Witten pointed out that
the partition function of twisted ${\cal N}=4$ super Yang-Mills theory
can have remarkably simple form,
if $X$ is a K\"ahler 4-manifold and vanishing theorem holds \cite{vafa-witten}.
In this situation, its partition function
is given by certain linear combination of
the summation of the Euler number of the moduli space of the ASD equations.
More precisely,  for twisted ${\cal N}=4~ {\cal G}/\Gamma_{\cal G}$ theory on $X$
with 't Hooft flux $v\in H^2(X,\Gamma_{\cal G})$,
the partition function of this theory is given by
\begin{equation}
Z^X_v(\tau):= q^{-{\frac{(r+1)\chi(X)}{24}}}\sum_k \chi({\cal M}(v,k))q^k
\;\;\;(q:=\exp(2\pi i \tau)),\label{zxvt}
\end{equation}
where ${\cal M}(v,k) $ is the moduli space of ASD connections
associated to ${\cal G}/\Gamma_{\cal G}$-principal bundle with 't Hooft flux $v$ and fractional instanton number $k\in \frac{1}{2|\Gamma_{\cal G}|}{\bf Z}$.
In (\ref{zxvt}), $\tau$ is the gauge coupling constant including theta angle,
$\chi(X)$ is Euler number of $X$ and $r$ is the rank of ${\cal G} $.
 The factor $q^{-{\frac{(r+1)\chi(X)}{24}}} $ is required by modular property like Dedekind's $\eta$ function.
On $K3$, $\chi(K3)=24$ and the partition function has the leading term $q^{-(r+1)}$.

With this result, Vafa and Witten
conjectured the behavior of the partition functions under the action of $SL(2,{\bf Z})$ on $\tau$. They started with 't Hooft's work \cite{tHooft}, where it was shown that the path integral with ${\bf Z}_N$-valued electric flux and
that with magnetic flux
are related by Fourier transform action.
Vafa and Witten combined the conjecture of strong/weak duality with this
't Hooft's result. Their conjecture is summarized by the following formula:

\begin{equation}
Z^X_v\left(-\frac{1}{\tau}\right)=\vert\Gamma_{\cal G}\vert^{-\frac{b_2(X)}{2}}
\left(
\frac{\tau}{i}
\right)^{-\frac{\chi(X)}{2}}
\sum_u\zeta_{|\Gamma_{\cal G}|}^{u\cdot v}Z_u^X(\tau),
\label{m-o}
\end{equation}
where $\zeta_{|\Gamma_{\cal G}|}=\exp(2\pi i/|\Gamma_{\cal G}|)$ \cite{vafa-witten}.

Hereafter we will concentrate on $X=K3$ case.
We discuss the gap condition on $K3$.
In $v=0$ case
${\cal M}(0,k)=:{\cal M}_k^{\cal G}  $ is  the moduli space of irreducible
ASD connections associated with ${\cal G}$-principal bundle with instanton number $k$.
Its dimension $\dim {\cal M}_k^{\cal G}$ is given by
Atiyah-Hitchin-Singer dimension formula \cite{AHS}:
 \beq
\dim{\cal M}^{\cal G}_k=4h({\cal G})(k-r)-4r,
\label{ahs}
\enq
where $h({\cal G})$ is the dual Coxeter number, $\dim{\cal G}$ is the dimension
of ${\cal G}$ and $k$ is the instanton number.
$ \dim{\cal M}^{\cal G}_k\ge 0$ stands for $k\ge r+1$.
That is, the moduli space of irreducible ASD connections with
$ADE$ gauge group ${\cal G}$ on $K3$ exist only for $k\ge r+1$ except for the $k=0$ case.
Thus the ${\cal G}$ partition function cannot have $q^{k-(r+1)}$ terms for $1\le k \le r$.
This condition is called gap condition \cite{vafa-witten, m-n, m-v, E-S, jin3}.

Finally we provide $SU(N)$ partition function on $K3$
in prime $N$ case satisfying all Vafa-Witten conjectures \cite{vafa-witten},
\beq
Z_{SU(N)}(\tau)=\frac{1}{N^3}\frac{1}{\eta(N\tau)^{24}}+\frac{1}{N^2}(\frac{1}{\eta(\frac{\tau}{N})^{24}}+\frac{1}{\eta(\frac{\tau+1}{N})^{24}}+\cdots+\frac{1}{\eta(\frac{\tau+N-1}{N})^{24}}).
\enq

\section{Our Previous Works}
\setcounter{equation}{0}

$K3$ has many special properties.
One is its orbifold construction.
$K3$ can be constructed by the following processes.
First we divide 4-torus $T^4$ by ${\bf Z}_2$ and obtain
quotient space $T^4/{\bf Z}_2=:S_0$.
Next we blow up its sixteen singularities by ${\cal O}(-2) $ curve
and obtain smooth K\"ahler surface $K3$.
$K3$ has the following data \cite{Fukaya}:
\beq
\chi(K3)=24, K_{K3}={\cal O},\mbox{Intersection form of~} K3~ \mbox{is~}(-E_8)^{\oplus 2}\oplus H^{\oplus 3}.
\enq
Here $K_{K3}$ is canonical bundle of $K3$ and ${\cal O}$ stands for trivial.
\[
H=\left(\begin{array}{cc}
0&1\\
1&0
\end{array}\right).
\]
We reproduced the above geometrical processes in the partition function
level for $SU(N)$ \cite{jin, jin2},where it is shown that the $SU(N)$ partition function on $K3$
has a sum of the product of two factors.
The first one is the contribution from $S_0$ and
is described by
Hecke transformation of order $N$ of $1/\eta^8(\tau)$.
The second one is the contribution
obtained from blowing up sixteen singularities
and is given by
the blow-up formula \cite{yoshioka, kap, vafa-witten}.
These two factors are summed up so that the resulting partition function
satisfies Montonen-Olive duality.
Our partition function is consistent to the fact that
$SU(N)$ partition function on $K3$ is the Hecke transformation of order $N$
of $1/\eta^{24}(\tau)$ \cite{vafa-witten, m-v, yoshihecke, lozano, mukai, nak}.
Key point of this verification is  that eta function $\eta(\tau/N) $
can be interpreted as blow-up formula
which is typically $\theta(\tau)/\eta^m(\tau)$ \cite{yoshioka, kap}.
Fortunately there are beautiful identities \cite{jin2}:
\beq
\frac{1}{\eta(\frac{\tau}{N})}=\frac{\theta_{A_{N-1}}(\tau)}{\eta^N(\tau)},
\label{moa}
\enq
which are already verified by celebrated denominator identity \cite{kac, mac}.

%$A\to ADE$
In the next work, we tried to determine the ${\cal N}=4~ ADE$ partition function
on $K3$ \cite{jin3}.
Using the generalization of (\ref{moa}) to the $ADE$ case,
we define the $ADE $ blow-up formula:
\beq
 \frac{\theta_{{\cal G}_r}(\tau)}{\eta^{r+1}(\tau)}=\mbox{eta product},
 \label{di}
\enq
where we note that these blow-up formula also have the form of eta product
and this fact is also verified by denominator identity \cite{kac, mac}.
To construct  ${\cal N}=4~ADE$ partition function on $K3$,
 which satisfies the duality conjecture,
we introduce 24-th power of (\ref{di}) and call it primary function.
First we generate a set of functions,
by modular transformation of the primary function
(adding $\tau \to \tau+1/m $ transformation in some cases).
Next we request Montonen-Olive duality to appropriate
liner combination of these functions.
Since it is difficult to produce the partition functions
directly from these functions,
we use the following observation.
By taking a subset of the functions,
appropriate liner combination of functions of this subset
has the same modular property
that a
piece of $SU(N)$
partition function has \cite{jin3}.
Here ${\cal G}_r$ and $SU(N)$ have the same Montonen-Olive duality.
We call a piece of $SU(N)$
partition functions as $G_j(\tau)$
and appropriate liner combination of functions of this subset as ${\tilde G}_j(\tau)$.
Then modular property of $G_j(\tau)$ and ${\tilde G}_j(\tau)$
is completely the same.
For ${\tilde G}_j(\tau)$, we use the same coefficient
as that of $G_j(\tau)$ in $Z_t(\tau)$ or $Z_{SU(N)/{\bf Z}_N}(\tau)$.
Finally we obtain the ${\cal N}=4~ADE$ partition function
satisfying Montonen-Olive duality.

Here
we explain why it is reliable to adopt $ADE$ blow-up formula
to construct ${\cal N}=4 ~ADE$ partition on $K3$.
There is a stringy picture:

\begin{tabular}{ccc}
\\
IIA on $K3\times T^2 \times ALE_{\small ADE}$ &$\leftrightarrow$& Hetero on $T^4\times T^2 \times ALE_{ADE}$\\
$\downarrow$ & &$\downarrow$\\
${\cal N}=4 ~ADE$ on $K3$ &$\leftrightarrow$& (${\cal N}=4~ U(1)$ on $ALE_{ADE})^{\otimes 24}$
\\
\\
\end{tabular}\\
%\[
%Z_{{\cal G}_r}(\tau)\sim \left( \frac{\theta_{{\cal G}_r}(\tau)}{\eta^{r+1}(\tau)}\right)^{24}
%\]\\
First line is IIA/Hetero duality \cite{hj}. On the IIA side, by compactifying $T^2\times ALE_{ADE}$, we obtain ${\cal N}=4 ~ADE$ on $K3$.
On the Hetero side, by compactifying $T^4\times T^2$,
we obtain $({\cal N}=4~U(1)$ on $ALE_{ADE})^{\otimes 24}$ \cite{jin3}.
The origin of ${\otimes 24}$ is explained by
the compactification of $K3\times T^2$
in IIA side.
Thus each side of
the second line is equivalent and their partition functions
in both side are the same.
On the other hand, the partition function of ${\cal N}=4~U(1)$ on $ALE_{ADE}$
was already obtained by Nakajima \cite{naka} and his
 results are very similar to ours in (\ref{di}).
Thus we adopt  24-th power of (\ref{di}) as a piece of
${\cal N}=4~ADE$ partition function on $K3$.

Unfortunately this approach has serious problem.
In fact, these partition functions for $D,E$
do not satisfy the gap condition.
To resolve this problem,
we took another approach.
In \cite{sasaki1} we invent the deformed Hecke operator,
which is the deformed version of Hecke operator with the dependence of the 't Hooft flux.
This operator satisfies the Montonen-Olive duality automatically.
Thus we only had to find the function to operate
so that $D,E$ partition function satisfies the gap condition.
At last, we could determine the $D,E$ partition functions,
which satisfy the Montonen-Olive duality and the gap condition
at the same time.
However we could not cover the all $ADE$ gauge groups.
In \cite{sasaki2} we considered the self-dualized $D,E$ partition function, which consists of $D,E$ part and $U(1)$ part.
We also determined the self-dualized $D,E$ partition function
so that this function satisfy the gap condition,
but is unsuccessful in covering the all $D,E$ gauge groups in the same way as in \cite{sasaki1}.

To cover the all $D,E$ gauge groups and construct general formulas
of all gauge groups, we will return to the approach of $ADE$ blow-up formula. Of course we will have a slight difference from the case in \cite{jin3}. In the next section we will propose	another approach of $ADE$ blow-up formula.

\section{$A_{N-1}$ Blow-up Formula}
\setcounter{equation}{0}
In \cite{jin2} we found the $A_{N-1}$ blow-up formula
\beq
\frac{1}{\eta(\frac{\tau}{N})}=\frac{\theta_{A_{N-1}}(\tau)}{\eta(\tau)^N},
\enq
\beq
\theta_{A_{N-1}}(\tau)=\sum_{m\in{\bf Z}^{N-1}}q^{\frac{1}{2}{}^t(m+\frac{\rho}{N})A_{N-1}(m+\frac{\rho}{N})},
\enq
where $A_{N-1}$ is Cartan Matrix and $\rho$ is the half sum of positive roots \cite{kac,mac,jin3,wakimoto1}.
In \cite{jin3} we obtained the analogue for $D,E$ case by generalizing (4.1) and produced $D,E$ partition function on $K3$ by using this so that the partition function satisfies Montonen-Olive duality.
Although theta function of (4.2) itself has	unfamiliar expression,
it might be written
 by familiar theta functions as follows:
\begin{conj} For odd prime $N$,
\beq
\frac{1}{\eta(\frac{\tau}{N})}=\frac{\theta_{A_{N-1}}(\tau)}{\eta(\tau)^N},
\enq
\beq
\theta_{A_{N-1}}(\tau)=\sum_{\beta\in \frac{M^*}{NM}}n_\beta\theta_{A_{N-1}}^\beta(\tau),{}^\exists n_\beta\in {\bf Z},
\enq
where $\theta_{A_{N-1}}^\beta(\tau)$ is a level $N~A_{N-1}$theta function with weight vector $\beta\in \frac{M^*}{NM}$. $M,M^*$ are root lattice and dual lattice respectively \cite{kac, wakimoto1}.
\end{conj}
Let us verify this conjecture for small rank.
Because we only need the function in $\tau$, we set $t=0,z=1$ for theta functions in \cite{kac, wakimoto1}.

\subsection{Proof of Conjecture 1 for Small Rank}
To verify the conjecture 1, we introduce a set of theta functions $\{ \Theta_{A_{N-1}}^{d,b}(\tau)\}$:
\beq
\frac{1}{\eta(\frac{a\tau+b}{d})}=:\frac{\Theta_{A_{N-1}}^{d,b}(\tau)}{\eta(\tau)^N},(a,b,d\in {\bf Z},ad=N,b<d).
\enq
By using $\Theta_{A_{N-1}}^{d,b}(\tau)$, we conjecture again
\begin{conj}
For odd prime $N$,
\beq
\Theta_{A_{N-1}}^{d,b}(\tau)=\sum_{\beta\in \frac{M^*}{NM}}n_\beta^{d,b}\zeta_{24}^{k_1}\zeta_{N^2}^{k_2}\theta_{A_{N-1}}^\beta(\tau),{}^\exists n_\beta^{d,b}\in {\bf Z},{}^\exists k_i\in{\bf Z},
\enq
where $\zeta_n=\exp(2\pi i/n)$.
\end{conj}
To prove the identity
between two modular forms,
one only needs to check the following two properties:
$q$-expansion and modular property \cite{wakimoto1}.
For $N=3,5,7$ the first step has already done in \cite{jin3}.
We will verify the second step mainly. For this purpose we cite
the modular property of theta functions \cite{kac, wakimoto1}.
\begin{thm}[Kac \cite{kac}]
For level $l$ and rank $r$ ${\cal G}_r$,
\beq
\theta_{{\cal G}_r}^\beta(-\frac{1}{\tau})=\left|\frac{M^*}{lM}\right|^{-\frac{1}{2}}(-i\tau)^{\frac{r}{2}}\sum_{\gamma\in \frac{M^*}{lM}}
\exp(-\frac{2\pi i\beta\cdot\gamma}{l})\theta_{{\cal G}_r}^\gamma(\tau),
\enq
where $\left|\frac{M^*}{lM}\right|$ is the number of elements of $\frac{M^*}{lM}$.
\end{thm}
In the case of $l=N,{\cal G}_r=A_{N-1}$,$\left|\frac{M^*}{lM}\right|=N^N$.
To treat this formula more easily, we decompose $N^N$ weight vectors
$\{\beta\in \frac{M^*}{NM}\}$ into each Weyl orbit as follows:
\beq
\{\beta\}=\{w(\lambda_0)\}\oplus\{w(\lambda_1)\}\oplus \cdots \oplus \{w(\lambda_m)\},
\enq
where $w$ is Weyl transformations \cite{kac, wakimoto1}.
It is well-known that there are independent orbits less than $N^N$ weight vectors for level $N~A_{N-1}$ theta functions.
 For each orbit $\{w(\lambda_j)\}$ one can investigate the modular property and verify the second step. Here we show the $N=3,5$ case explicitly.
Before moving to explicit verification, we introduce the following notation for $\theta_{A_{N-1}}^\beta(\tau)$:
\beq
\theta_{A_{N-1}}^{\beta}(\tau):=\sum_{m\in{\bf Z}^{N-1}}q^{\frac{N}{2}{}^t(m+\frac{1}{N}A_{N-1}^{-1}\beta)A_{N-1}(m+\frac{1}{N}A_{N-1}^{-1}\beta)},{}^t\beta=(b_1,b_2,\ldots,b_{N-1}),b_i\in {\bf Z}.
\enq
\paragraph{$A_2$}~\\
For level $3~A_2$ theta functions, there are 6 Weyl orbits(see appendix).
By comparing $q$-expansion of these orbits with that of $\Theta_{A_2}^{d,b}(\tau)$, we can introduce
\beq
 {\tilde \Theta}_{A_2}^{1,0}(\tau):=\theta_{A_2}^{\lambda_0}-\theta_{A_2}^{\lambda_1^3},
\enq
\beq
{\tilde \Theta}_{A_2}^{3,b}(\tau):=\zeta_{24}^{-b}\zeta_9^b(\theta_{A_2}^{\lambda_1^1}+\zeta_3^b\theta_{A_2}^{\lambda_1^2}+\zeta_3^{2b}\theta_{A_2}^{\lambda_1^4}),b=0,1,2,
\enq
where $\zeta_n=\exp(\frac{2\pi i}{n})$.
We introduce ${\tilde\Theta}_{A_2}^{d,b}(\tau)$ whose $q$-expansion is the same as that of $\Theta_{A_2}^{d,b}(\tau)$.
On the other hands we can also investigate modular property of these theta functions by using (4.6). We obtain
\beq
{\tilde \Theta}_{A_2}^{1,0}(-\frac{1}{\tau})=(-i\tau)\sqrt{3}{\tilde \Theta}_{A_2}^{3,0}(\tau),
\enq
\beq
{\tilde \Theta}_{A_2}^{3,1}(-\frac{1}{\tau})=(-i\tau)\zeta_{24}{\tilde \Theta}_{A_2}^{3,2}(\tau).
\enq
This modular property is the same as that of $\Theta_{A_2}^{d,b}(\tau)$.
Thus we conclude the equivalence between $\Theta_{A_2}^{d,b}(\tau)$ and ${\tilde\Theta}_{A_2}^{d,b}(\tau)$ completely.
\paragraph{$A_4$}~\\
For level $5~A_4$ theta functions, there are 66 Weyl orbits.
By comparing $q$-expansion of these orbits with that of $\Theta_{A_4}^{d,b}(\tau)$, we can introduce
\beq
 {\tilde \Theta}_{A_4}^{1,0}(\tau):=\theta_{A_4}^{\lambda_0}+2\theta_{A_4}^{\lambda_1^5}+2\theta_{A_4}^{\lambda_1^{10}}-5\theta_{A_4}^{\lambda_9},
\enq
\beqy
{\tilde \Theta}_{A_4}^{5,b}(\tau)&:=&\zeta_{24}^{-b}(5\theta_{A_4}^{\lambda_9}
\no\\
&&
+\zeta_5^b(\theta_{A_4}^{\lambda_6^1}+2\theta_{A_4}^{\lambda_{10}^1}+2\theta_{A_4}^{\lambda_{11}^2})
\no\\
&&
+\zeta_5^{2b}(\theta_{A_4}^{\lambda_8^2}+2\theta_{A_4}^{\lambda_{7}^2}+2\theta_{A_4}^{\lambda_{12}^2})
\no\\
&&
+\zeta_5^{3b}(\theta_{A_4}^{\lambda_8^1}+2\theta_{A_4}^{\lambda_{7}^1}+2\theta_{A_4}^{\lambda_{12}^1})
\no\\
&&
+\zeta_5^{4b}(\theta_{A_4}^{\lambda_6^2}+2\theta_{A_4}^{\lambda_{10}^2}+2\theta_{A_4}^{\lambda_{11}^1})),b=0,1,2,3,4.
\enqy
We introduce ${\tilde\Theta}_{A_4}^{d,b}(\tau)$ whose $q$-expansion is the same as that of $\Theta_{A_4}^{d,b}(\tau)$.
On the other hands we can also investigate modular property of these theta functions by using (4.6). We obtain
\beq
{\tilde \Theta}_{A_4}^{1,0}(-\frac{1}{\tau})=(-i\tau)^2\sqrt{5}{\tilde \Theta}_{A_4}^{5,0}(\tau),
\enq
\beq
{\tilde \Theta}_{A_4}^{5,1}(-\frac{1}{\tau})=(-i\tau)^2\zeta_{8}{\tilde \Theta}_{A_4}^{5,4}(\tau),
\enq
\beq
{\tilde \Theta}_{A_4}^{5,b}(-\frac{1}{\tau})=(-i\tau)^2{\tilde \Theta}_{A_4}^{5,b}(\tau),b=2,3.
\enq
This modular property is the same as that of $\Theta_{A_4}^{d,b}(\tau)$.
Thus we conclude the equivalence between $\Theta_{A_4}^{d,b}(\tau)$ and ${\tilde\Theta}_{A_4}^{d,b}(\tau)$ completely.

\subsection{Even $N$ case}
Unfortunately the conjecture 1(2) cannot be  valid in the even $N$ case because level $N~A_{N-1}$ theta functions do not have the corresponding $q$-expansion.
Thus we generalize the conjecture 2 to
\begin{conj}
\beq
(\Theta_{A_{N-1}}^{d,b}(\tau))^m=\sum_{\{\beta\}\in (\frac{M^*}{NM})^m}n_{\{\beta\}}^{d,b}\zeta_{24}^{k_1}\zeta_{2N^2}^{k_2}\theta_{A_{N-1}}^{\beta_1}(\tau)\theta_{A_{N-1}}^{\beta_2}(\tau)\cdots\theta_{A_{N-1}}^{\beta_m}(\tau),{}^\exists n_{\{\beta\}}^{d,b}\in {\bf Z},{}^\exists k_i\in{\bf Z},{}^\exists m\in{\bf Z}.
\enq
\end{conj}
We have done the verification of conjecture 3 of $q$-expansion level in $N=2,4,6$.
We will show that of modular property in $N=2,4$ in the same way as odd prime $N$ case.

\paragraph{$A_1$}~\\
For level $2~A_1$ theta functions, there are 3 Weyl orbits:
\beqy
{}^t\lambda_0&=&0,\no\\
{}^t\lambda_1^k&=&k,k=1,2.
\enqy
By comparing $q$-expansion of these orbits with that of $(\Theta_{A_1}^{d,b}(\tau))^2$, we can introduce
\beq
({\tilde \Theta}_{A_1}^{1,0}(\tau))^2:=\theta_{A_1}^{\lambda_1^1}(\theta_{A_1}^{\lambda_0}+\theta_{A_1}^{\lambda_1^2}),
\enq
\beq
 ({\tilde \Theta}_{A_1}^{2,b}(\tau))^2:=\frac{1}{2}\zeta_6^{-b}\zeta_4^b\theta_{A_1}^{\lambda_1^1}(\theta_{A_1}^{\lambda_0}+(-1)^b\theta_{A_1}^{\lambda_1^2}),b=0,1.
\enq
We define $({\tilde\Theta}_{A_1}^{d,b}(\tau))^2$ whose $q$-expansion is the same as that of $(\Theta_{A_1}^{d,b}(\tau))^2$.
Prefactor $\frac{1}{2}$ of (4.22) is a generalization of conjecture 3.
On the other hands we can also investigate modular property of these theta functions by using (4.6). We obtain
\beq
({\tilde \Theta}_{A_1}^{1,0}(-\frac{1}{\tau}))^2=(-i\tau)2({\tilde \Theta}_{A_1}^{2,0}(\tau))^2,
\enq
\beq
({\tilde \Theta}_{A_1}^{2,1}(-\frac{1}{\tau}))^2=(-i\tau)({\tilde \Theta}_{A_1}^{2,1}(\tau))^2.
\enq
This modular property is the same as that of $(\Theta_{A_1}^{d,b}(\tau))^2$.
Thus we conclude the equivalence between $(\Theta_{A_1}^{d,b}(\tau))^2$ and $({\tilde\Theta}_{A_1}^{d,b}(\tau))^2$ completely.
\paragraph{$A_3$}~\\
For level $4~A_3$ theta functions, there are 22 Weyl orbits.
By comparing $q$-expansion of these orbits with that of $(\Theta_{A_3}^{d,b}(\tau))^m$,
we find that $m=3$ is the least case of conjecture 3.
We introduce
\beqy
&&({\tilde \Theta}_{A_3}^{1,0}(\tau))^3
\no\\&&
:=(\theta_{A_3}^{\lambda_0})^3
+8(\theta_{A_3}^{\lambda_4^2})^3
-(\theta_{A_3}^{\lambda_1^8})^3
+(\theta_{A_3}^{\lambda_0})^2\theta_{A_3}^{\lambda_1^8}
+24\theta_{A_3}^{\lambda_0}(\theta_{A_3}^{\lambda_4^2})^2
-\theta_{A_3}^{\lambda_0}(\theta_{A_3}^{\lambda_1^8})^2
\no\\
&&
-24\theta_{A_3}^{\lambda_4^2}(\theta_{A_3}^{\lambda_2^2})^2
-8\theta_{A_3}^{\lambda_0}(\theta_{A_3}^{\lambda_1^4})^2
-12\theta_{A_3}^{\lambda_0}(\theta_{A_3}^{\lambda_2^2})^2
+12\theta_{A_3}^{\lambda_1^8}(\theta_{A_3}^{\lambda_2^2})^2
%s0^3+(8)*s2d^3+(-1)*s8a^3+(1)*s0^2*s8a+(24)*s0*s2d^2+(-1)*s0*s8a^2
%+((-24)*s2d*s2b^2+(-8)*s0*s4a^2+(-12)*s0*s2b^2+(12)*s8a*s2b^2))
,
\enqy
\beqy
&&({\tilde \Theta}_{A_3}^{2,0}(\tau))^3
\no\\&&
:=(\theta_{A_3}^{\lambda_0})^2\theta_{A_3}^{\lambda_1^2}
-12(\theta_{A_3}^{\lambda_0})^2\theta_{A_3}^{\lambda_1^6}
-12(\theta_{A_3}^{\lambda_4^2})^2\theta_{A_3}^{\lambda_1^2}
-6\theta_{A_3}^{\lambda_7}(\theta_{A_3}^{\lambda_2^3})^2
-10(\theta_{A_3}^{\lambda_2^1})^2\theta_{A_3}^{\lambda_2^3}
+2(\theta_{A_3}^{\lambda_2^3})^2\theta_{A_3}^{\lambda_2^1}
\no\\&&
+20\theta_{A_3}^{\lambda_7}\theta_{A_3}^{\lambda_2^1}\theta_{A_3}^{\lambda_2^3}
+36\theta_{A_3}^{\lambda_7}\theta_{A_3}^{\lambda_4^2}\theta_{A_3}^{\lambda_8}
+13\theta_{A_3}^{\lambda_0}\theta_{A_3}^{\lambda_2^3}\theta_{A_3}^{\lambda_4^1}
-\theta_{A_3}^{\lambda_0}\theta_{A_3}^{\lambda_4^2}\theta_{A_3}^{\lambda_1^6}
+12\theta_{A_3}^{\lambda_0}\theta_{A_3}^{\lambda_2^1}\theta_{A_3}^{\lambda_8}
-37\theta_{A_3}^{\lambda_1^8}\theta_{A_3}^{\lambda_7}\theta_{A_3}^{\lambda_4^1}
\no\\&&
+\theta_{A_3}^{\lambda_1^8}\theta_{A_3}^{\lambda_2^1}\theta_{A_3}^{\lambda_8}
+\theta_{A_3}^{\lambda_1^8}\theta_{A_3}^{\lambda_2^3}\theta_{A_3}^{\lambda_8}
+(\theta_{A_3}^{\lambda_7})^3
%(1)*s0^2*s2a+(-12)*s0^2*s6a+(-12)*s2d^2*s2a+(-6)*sg*s3b^2+(-10)*sb^2*%s3b+(2)*s3b^2*sb+(20)*sg*sb*s3b+(36)*sg*s2d*sh+(13)*s0*s3b*sd+(-1)*s0%*s2d*s6a+(12)*s0*sb*sh+(-37)*s8a*sg*sd+(1)*s8a*sb*sh+(1)*s8a*s3b*sh+(%-8)*sg^3
,
\enqy
\beqy
&&({\tilde \Theta}_{A_3}^{2,1}(\tau))^3
\no\\&&
:=\zeta_{16}^{-1}(
-8\theta_{A_3}^{\lambda_2^1}(\theta_{A_3}^{\lambda_7})^2
+64(\theta_{A_3}^{\lambda_0})^2\theta_{A_3}^{\lambda_1^2}
-16\theta_{A_3}^{\lambda_7}(\theta_{A_3}^{\lambda_2^3})^2
-14(\theta_{A_3}^{\lambda_2^1})^2\theta_{A_3}^{\lambda_2^3}
-82(\theta_{A_3}^{\lambda_2^3})^2\theta_{A_3}^{\lambda_2^1}
+80\theta_{A_3}^{\lambda_1^8}\theta_{A_3}^{\lambda_2^3}\theta_{A_3}^{\lambda_4^1}
\no\\&&
+24\theta_{A_3}^{\lambda_7}\theta_{A_3}^{\lambda_2^1}\theta_{A_3}^{\lambda_2^3}
-63\theta_{A_3}^{\lambda_0}\theta_{A_3}^{\lambda_2^1}\theta_{A_3}^{\lambda_4^1}
+8\theta_{A_3}^{\lambda_7}\theta_{A_3}^{\lambda_4^2}\theta_{A_3}^{\lambda_8}
-\theta_{A_3}^{\lambda_0}\theta_{A_3}^{\lambda_2^3}\theta_{A_3}^{\lambda_4^1}
+16\theta_{A_3}^{\lambda_1^8}\theta_{A_3}^{\lambda_2^1}\theta_{A_3}^{\lambda_4^1}
+\theta_{A_3}^{\lambda_1^8}\theta_{A_3}^{\lambda_2^1}\theta_{A_3}^{\lambda_8}
\no\\&&
-16\theta_{A_3}^{\lambda_1^8}\theta_{A_3}^{\lambda_7}\theta_{A_3}^{\lambda_8}
-\theta_{A_3}^{\lambda_1^8}\theta_{A_3}^{\lambda_2^3}\theta_{A_3}^{\lambda_8}
+(\theta_{A_3}^{\lambda_7})^3)
%(-8)*sb*sg^2+(64)*s0^2*s2a+(-16)*sg*s3b^2+(-14)*sb^2*s3b+(-82)*s3b^2*%sb+(80)*s8a*s3b*sd+(24)*sg*sb*s3b+(-63)*s0*sb*sd+(8)*sg*s2d*sh+(-1)*s%0*s3b*sd+(16)*s8a*sb*sd+(8)*s8a*sg*sd+(1)*s8a*sb*sh+(-16)*s8a*sg*sh+(%-1)*s8a*s3b*sh
,
\enqy
\beqy
&&({\tilde \Theta}_{A_3}^{4,b}(\tau))^3
\no\\&&
:=(-1)^b\zeta_{32}^{15b}((
-10\theta_{A_3}^{\lambda_3^1}\theta_{A_3}^{\lambda_0}\theta_{A_3}^{\lambda_7}
+\theta_{A_3}^{\lambda_2^1}\theta_{A_3}^{\lambda_4^1}\theta_{A_3}^{\lambda_1^1}
+34\theta_{A_3}^{\lambda_7}\theta_{A_3}^{\lambda_8}\theta_{A_3}^{\lambda_1^1}
+2\theta_{A_3}^{\lambda_2^1}\theta_{A_3}^{\lambda_8}\theta_{A_3}^{\lambda_1^1}
+10\theta_{A_3}^{\lambda_7}\theta_{A_3}^{\lambda_1^5}\theta_{A_3}^{\lambda_4^2}
+\theta_{A_3}^{\lambda_2^3}\theta_{A_3}^{\lambda_8}\theta_{A_3}^{\lambda_1^1}
\no\\&&
+58\theta_{A_3}^{\lambda_6}\theta_{A_3}^{\lambda_5}\theta_{A_3}^{\lambda_3^3}
-47\theta_{A_3}^{\lambda_5}\theta_{A_3}^{\lambda_1^7}\theta_{A_3}^{\lambda_1^8}
+3\theta_{A_3}^{\lambda_2^2}\theta_{A_3}^{\lambda_1^3}\theta_{A_3}^{\lambda_2^1}
+3\theta_{A_3}^{\lambda_2^2}\theta_{A_3}^{\lambda_1^3}\theta_{A_3}^{\lambda_8}
+24\theta_{A_3}^{\lambda_6}\theta_{A_3}^{\lambda_1^2}\theta_{A_3}^{\lambda_1^5}
-32\theta_{A_3}^{\lambda_6}\theta_{A_3}^{\lambda_3^5}\theta_{A_3}^{\lambda_7}
\no\\&&
+43\theta_{A_3}^{\lambda_6}\theta_{A_3}^{\lambda_3^5}\theta_{A_3}^{\lambda_2^1}
+13\theta_{A_3}^{\lambda_6}\theta_{A_3}^{\lambda_3^5}\theta_{A_3}^{\lambda_2^3}
+4\theta_{A_3}^{\lambda_6}\theta_{A_3}^{\lambda_1^7}\theta_{A_3}^{\lambda_7}
-44\theta_{A_3}^{\lambda_1^4}\theta_{A_3}^{\lambda_3^7}\theta_{A_3}^{\lambda_2^1}
+9\theta_{A_3}^{\lambda_5}\theta_{A_3}^{\lambda_3^7}\theta_{A_3}^{\lambda_0}
+56\theta_{A_3}^{\lambda_5}\theta_{A_3}^{\lambda_1^7}\theta_{A_3}^{\lambda_8}
)
%(-10)*tc*t0*tg+(1)*tb*td*ta+(34)*tg*th*ta+(2)*tb*th*ta+(10)*tg*t5a*t2%d+(1)*t3b*th*ta+q*((58)*tf*te*t3c+(-47)*te*t7c*t8a+(3)*t2b*t3a*tb+(3)%*t2b*t3a*t3b+(24)*tf*t2a*t5a+(-32)*tf*t5c*tg+(43)*tf*t5c*tb+(13)*tf*t%5c*t3b+(4)*tf*t7a*tg+(-44)*t4a*t7c*tb+(9)*te*t7c*t0+(56)*te*t7a*th)
\no\\&&
+\zeta_4^b(62\theta_{A_3}^{\lambda_1^7}\theta_{A_3}^{\lambda_0}\theta_{A_3}^{\lambda_2^3}
-172\theta_{A_3}^{\lambda_1^2}\theta_{A_3}^{\lambda_3^1}\theta_{A_3}^{\lambda_0}
+\theta_{A_3}^{\lambda_2^3}\theta_{A_3}^{\lambda_1^5}\theta_{A_3}^{\lambda_4^1}
-319\theta_{A_3}^{\lambda_2^3}\theta_{A_3}^{\lambda_8}\theta_{A_3}^{\lambda_1^5}
+252\theta_{A_3}^{\lambda_1^6}\theta_{A_3}^{\lambda_1^8}\theta_{A_3}^{\lambda_1^5}
+79\theta_{A_3}^{\lambda_1^6}\theta_{A_3}^{\lambda_4^2}\theta_{A_3}^{\lambda_1^5}
\no\\&&
+71\theta_{A_3}^{\lambda_5}\theta_{A_3}^{\lambda_1^8}\theta_{A_3}^{\lambda_3^3}
+6\theta_{A_3}^{\lambda_5}\theta_{A_3}^{\lambda_0}\theta_{A_3}^{\lambda_1^1}
+103\theta_{A_3}^{\lambda_1^2}\theta_{A_3}^{\lambda_1^1}\theta_{A_3}^{\lambda_4^1}
+\theta_{A_3}^{\lambda_1^2}\theta_{A_3}^{\lambda_1^8}\theta_{A_3}^{\lambda_1^5}
+204\theta_{A_3}^{\lambda_1^4}\theta_{A_3}^{\lambda_3^3}\theta_{A_3}^{\lambda_2^3}
+261\theta_{A_3}^{\lambda_2^2}\theta_{A_3}^{\lambda_3^3}\theta_{A_3}^{\lambda_2^3}
\no\\&&
+66\theta_{A_3}^{\lambda_3^5}\theta_{A_3}^{\lambda_0}\theta_{A_3}^{\lambda_2^1}
-143\theta_{A_3}^{\lambda_3^5}\theta_{A_3}^{\lambda_1^8}\theta_{A_3}^{\lambda_7}
+3\theta_{A_3}^{\lambda_3^5}\theta_{A_3}^{\lambda_1^8}\theta_{A_3}^{\lambda_2^2}
+103\theta_{A_3}^{\lambda_1^6}\theta_{A_3}^{\lambda_3^3}\theta_{A_3}^{\lambda_8}
-122\theta_{A_3}^{\lambda_1^4}\theta_{A_3}^{\lambda_3^3}\theta_{A_3}^{\lambda_7}
-212\theta_{A_3}^{\lambda_1^7}\theta_{A_3}^{\lambda_4^2}\theta_{A_3}^{\lambda_2^3}
\no\\&&
-62\theta_{A_3}^{\lambda_1^7}\theta_{A_3}^{\lambda_1^8}\theta_{A_3}^{\lambda_2^1}
+126\theta_{A_3}^{\lambda_6}\theta_{A_3}^{\lambda_1^3}\theta_{A_3}^{\lambda_7}
+134\theta_{A_3}^{\lambda_6}\theta_{A_3}^{\lambda_1^7}\theta_{A_3}^{\lambda_1^6}
-314\theta_{A_3}^{\lambda_6}\theta_{A_3}^{\lambda_1^3}\theta_{A_3}^{\lambda_2^3}
)
%(62)*t7a*t0*t3b+(-172)*t2a*tc*t0+(1)*t3b*t5a*td+(-319)*t3b*th*t5a+(25%2)*t6a*t8a*t5a+(79)*t6a*t2d*t5a+(71)*te*t8a*t3c+(6)*te*t0*ta+(103)*t2%a*ta*td+(1)*t2a*t8a*t5a+(204)*t4a*t3c*t3b+(261)*t2b*t3c*t3b+(66)*t5c*%t0*tb+(-143)*t5c*t8a*tg+(3)*t5c*t8a*t3b+(103)*t6a*t3c*th+(-122)*t4a*t%3c*tg+(-212)*t7a*t2d*t3b+(-62)*t7a*t8a*tb+q*((126)*tf*t3a*tg+(134)*tf%*t7a*t6a+(-314)*tf*t3a*t3b)
\no\\&&
+\zeta_4^{2b}(-62\theta_{A_3}^{\lambda_5}\theta_{A_3}^{\lambda_1^1}\theta_{A_3}^{\lambda_8}
+\theta_{A_3}^{\lambda_1^2}\theta_{A_3}^{\lambda_8}\theta_{A_3}^{\lambda_1^5}
-15\theta_{A_3}^{\lambda_1^7}\theta_{A_3}^{\lambda_1^7}\theta_{A_3}^{\lambda_8}
+3\theta_{A_3}^{\lambda_3^5}\theta_{A_3}^{\lambda_2^3}\theta_{A_3}^{\lambda_8}
+3\theta_{A_3}^{\lambda_2^2}\theta_{A_3}^{\lambda_2^3}\theta_{A_3}^{\lambda_1^5}
+2\theta_{A_3}^{\lambda_3^7}\theta_{A_3}^{\lambda_1^8}\theta_{A_3}^{\lambda_2^1}
\no\\&&
-11\theta_{A_3}^{\lambda_3^7}\theta_{A_3}^{\lambda_0}\theta_{A_3}^{\lambda_7}
+41\theta_{A_3}^{\lambda_2^2}\theta_{A_3}^{\lambda_3^1}\theta_{A_3}^{\lambda_2^1}
+106\theta_{A_3}^{\lambda_2^2}\theta_{A_3}^{\lambda_3^1}\theta_{A_3}^{\lambda_7}
+68\theta_{A_3}^{\lambda_2^2}\theta_{A_3}^{\lambda_3^1}\theta_{A_3}^{\lambda_2^3}
-24\theta_{A_3}^{\lambda_3^1}\theta_{A_3}^{\lambda_1^2}\theta_{A_3}^{\lambda_4^1}
-23\theta_{A_3}^{\lambda_3^1}\theta_{A_3}^{\lambda_1^6}\theta_{A_3}^{\lambda_4^1}
\no\\&&
+4\theta_{A_3}^{\lambda_1^4}\theta_{A_3}^{\lambda_1^2}\theta_{A_3}^{\lambda_1^1}
+12\theta_{A_3}^{\lambda_1^4}\theta_{A_3}^{\lambda_1^2}\theta_{A_3}^{\lambda_3^3}
-8\theta_{A_3}^{\lambda_2^2}\theta_{A_3}^{\lambda_1^2}\theta_{A_3}^{\lambda_1^1}
+\theta_{A_3}^{\lambda_1^7}\theta_{A_3}^{\lambda_1^6}\theta_{A_3}^{\lambda_1^8}
+\theta_{A_3}^{\lambda_1^7}\theta_{A_3}^{\lambda_1^6}\theta_{A_3}^{\lambda_1^8}
+8\theta_{A_3}^{\lambda_5}\theta_{A_3}^{\lambda_6}\theta_{A_3}^{\lambda_3^5}
\no\\&&
-23\theta_{A_3}^{\lambda_1^3}\theta_{A_3}^{\lambda_2^2}\theta_{A_3}^{\lambda_5}
-10\theta_{A_3}^{\lambda_3^7}\theta_{A_3}^{\lambda_2^2}\theta_{A_3}^{\lambda_5}
+54\theta_{A_3}^{\lambda_3^7}\theta_{A_3}^{\lambda_1^2}\theta_{A_3}^{\lambda_6})
%(-62)*te*ta*th+(1)*t2a*th*t5a+(-15)*t7a*tg*th+(3)*t5c*t3b*th+(3)*t2b*%t3b*t5a+(2)*t7c*t8a*tb+(-11)*t7c*t0*tg+(41)*t2b*tc*tb+(106)*t2b*tc*tg%+(68)*t2b*tc*t3b+(-24)*tc*t2a*td+(-23)*tc*t6a*td+(4)*t4a*t2a*ta+(12)*%t4a*t2a*t3c+(-8)*t2b*t2a*ta+(1)*t7a*t2a*t0+(1)*t7a*t6a*t8a+q*((8)*te*%tf*t5c+(-23)*t3a*t2b*te+(-10)*t7c*t2b*te+(54)*t7c*t2a*tf)
\no\\&&
+\zeta_4^{3b}(-14\theta_{A_3}^{\lambda_1^7}\theta_{A_3}^{\lambda_1^6}\theta_{A_3}^{\lambda_4^1}
-16\theta_{A_3}^{\lambda_3^5}\theta_{A_3}^{\lambda_1^4}\theta_{A_3}^{\lambda_2^1}
+58\theta_{A_3}^{\lambda_3^5}\theta_{A_3}^{\lambda_1^2}\theta_{A_3}^{\lambda_4^1}
+44\theta_{A_3}^{\lambda_1^7}\theta_{A_3}^{\lambda_2^2}\theta_{A_3}^{\lambda_2^1}
-5\theta_{A_3}^{\lambda_3^5}\theta_{A_3}^{\lambda_2^2}\theta_{A_3}^{\lambda_2^1}
+3\theta_{A_3}^{\lambda_3^7}\theta_{A_3}^{\lambda_2^3}\theta_{A_3}^{\lambda_8}
\no\\&&
-63\theta_{A_3}^{\lambda_3^7}\theta_{A_3}^{\lambda_2^1}\theta_{A_3}^{\lambda_4^1}
+62\theta_{A_3}^{\lambda_1^6}\theta_{A_3}^{\lambda_6}\theta_{A_3}^{\lambda_3^3}
+19\theta_{A_3}^{\lambda_1^4}\theta_{A_3}^{\lambda_5}\theta_{A_3}^{\lambda_1^1}
+19\theta_{A_3}^{\lambda_2^2}\theta_{A_3}^{\lambda_5}\theta_{A_3}^{\lambda_1^1}
+2\theta_{A_3}^{\lambda_3^7}\theta_{A_3}^{\lambda_1^2}\theta_{A_3}^{\lambda_0}
-26\theta_{A_3}^{\lambda_3^1}\theta_{A_3}^{\lambda_6}\theta_{A_3}^{\lambda_7}
\no\\&&
+14\theta_{A_3}^{\lambda_6}\theta_{A_3}^{\lambda_2^1}\theta_{A_3}^{\lambda_1^5}
+4\theta_{A_3}^{\lambda_6}\theta_{A_3}^{\lambda_7}\theta_{A_3}^{\lambda_1^5}
+9\theta_{A_3}^{\lambda_3^5}\theta_{A_3}^{\lambda_5}\theta_{A_3}^{\lambda_0}
-24\theta_{A_3}^{\lambda_1^7}\theta_{A_3}^{\lambda_5}\theta_{A_3}^{\lambda_0}
+2\theta_{A_3}^{\lambda_1^3}\theta_{A_3}^{\lambda_1^2}\theta_{A_3}^{\lambda_0}
+\theta_{A_3}^{\lambda_1^3}\theta_{A_3}^{\lambda_1^2}\theta_{A_3}^{\lambda_1^8}
\no\\&&
+\theta_{A_3}^{\lambda_1^3}\theta_{A_3}^{\lambda_1^6}\theta_{A_3}^{\lambda_0}
+44\theta_{A_3}^{\lambda_1^7}\theta_{A_3}^{\lambda_1^2}\theta_{A_3}^{\lambda_4^1}
-6\theta_{A_3}^{\lambda_5}\theta_{A_3}^{\lambda_6}\theta_{A_3}^{\lambda_3^7}
))
%(-14)*t7a*t6a*td+(-16)*t5c*t4a*tb+(58)*t5c*t2a*td+(44)*t7a*t2b*tb+(-5%)*t5c*t2b*tb+(3)*t7c*t3b*th+(-63)*t7c*tb*td+(62)*t6a*tf*t3c+(19)*t4a*%te*ta+(19)*t2b*te*ta+(2)*t7c*t2a*t0+(-26)*tc*tf*tg+(14)*tf*tb*t5a+(4)%*tf*tg*t5a+(9)*t5c*te*t0+(-24)*t7a*te*t0+(2)*t3a*t2a*t0+(1)*t3a*t2a*t%8a+(1)*t3a*t6a*t0+(44)*t7a*t2a*td+q*((-6)*te*tf*t7c)
,b=0,1,2,3.
\enqy
We introduce $({\tilde\Theta}_{A_3}^{d,b}(\tau))^3$ whose $q$-expansion is the same as that of $(\Theta_{A_3}^{d,b}(\tau))^3$.
Note that the expression of (4.24)...(4.27) has an ambiguity because of the identities between theta functions.
On the other hands we can also investigate modular property of these theta functions by using (4.6). We obtain
\beq
({\tilde \Theta}_{A_3}^{1,0}(-\frac{1}{\tau}))^3=(-i\tau)^{\frac{9}{2}}2^3({\tilde \Theta}_{A_3}^{4,0}(\tau))^3,
\enq
\beq
({\tilde \Theta}_{A_3}^{2,0}(-\frac{1}{\tau}))^3=(-i\tau)^{\frac{9}{2}}({\tilde \Theta}_{A_3}^{2,0}(\tau))^3,
\enq
\beq
({\tilde \Theta}_{A_3}^{2,1}(-\frac{1}{\tau}))^3=(-i\tau)^{\frac{9}{2}}2\sqrt{2}({\tilde \Theta}_{A_3}^{4,2}(\tau))^3,
\enq
\beq
({\tilde \Theta}_{A_3}^{4,1}(-\frac{1}{\tau}))^3=(-i\tau)^{\frac{9}{2}}\zeta_{4}({\tilde \Theta}_{A_3}^{4,3}(\tau))^3.
\enq
This modular property is the same as that of $(\Theta_{A_3}^{d,b}(\tau))^3$.
Thus we conclude the equivalence between $(\Theta_{A_3}^{d,b}(\tau))^3$ and $({\tilde\Theta}_{A_3}^{d,b}(\tau))^3$ completely.
Similarly  we have checked $m=4$ case in $q$-expansion level.
Moreover we assume that $m=5$ case is also valid in consideration of leading $q$-terms.

In this section we described the theta function $\theta_{A_{N-1}}(\tau)$, which comes from the denominator identity \cite{mac,kac,jin2},
by level $N~A_{N-1}$ theta functions.
$\theta_{A_{N-1}}(\tau)$ itself has no direct connection with the $A_{N-1}$ character.
However we consider that level $N~A_{N-1}$ theta functions can have connection with the $A_{N-1}$ character.
Here in $A_1$ case we will describe the theta function $\theta_{A_{1}}(\tau)$ written by level $2~A_1$ theta functions in terms of  the $A_1$ character as follows. First we provide the $A_1$ character $\chi_{\lambda_1^1}(\tau)$ described by level $2~A_1$ theta functions and eta functions:
\beqy
\chi_{\lambda_1^1}(\tau)&=&\frac{\eta(2\tau)}{\eta(\tau)^2}\theta_{A_1}^{\lambda_1^1}(\tau)
\\
&=&
\frac{\theta_{A_1}^{\lambda_1^1}(\tau)}{\eta(\tau)^2}\eta(2\tau),
\enqy
where $\chi_{\lambda_1^1}(\tau)$ is the Weyl-Kac character formula associated to weight vector $\lambda_1^1$ \cite{kac, kac1, wakimoto1, wakimoto2}.
We remark that the factor $\frac{\eta(2\tau)}{\eta(\tau)^2}=c_{\lambda_1^1}^{\lambda_1^1}(\tau)$ in (4.33) is called the string function through $\lambda_1^1$. In general the character can be written by string functions and theta functions \cite{kac, wakimoto1}.
From (4.34) and the modular transformation of it, we obtain
\beq
\frac{\theta_{A_1}^{\lambda_1^1}(\tau)}{\eta(\tau)^2}
=\frac{\chi_{\lambda_1^1}(\tau)}{\eta(2\tau)},
\enq
\beq
\frac{\theta_{A_1}^{\lambda_0}(\tau)+\theta_{A_1}^{\lambda_1^2}(\tau)}{\eta(\tau)^2}
=\frac{\sqrt{2}(\chi_{\lambda_0}(\tau)+\chi_{\lambda_1^2}(\tau))}{\eta(\frac{\tau+1}{2})}.
\enq
Finally by combining (4.22) we obtain
\beq
\frac{1}{\eta(\frac{\tau}{2})^2}=\frac{\theta_{A_1}(\tau)^2}{\eta(\tau)^4}=\frac{\chi_{\lambda_1^1}(\tau)(\chi_{\lambda_0}(\tau)+\chi_{\lambda_1^2}(\tau))}{\sqrt{2}\eta(2\tau)\eta{(\frac{\tau+1}{2}})}.
\enq

\section{$D,E$ Case}
\setcounter{equation}{0}
In the previous section we considered $A_{N-1}$ blow-up formula
described by level $N~A_{N-1}$ theta functions.
We generalize this formula
to that for $ADE$ case described by level $|\Gamma_{\cal G}|~{\cal G}_r$ theta functions. Level $|\Gamma_{\cal G}|$ comes from Montonen-Olive duality associated with ${\cal N}=4$ super Yang-Mills for ${\cal G}_r$. More precisely we introduce an analogue of $A_{N-1}$ blow-up formula
in the previous section by combining
\beq
\frac{\theta_{{\cal G}_r}(\tau)}{\eta(\tau)^{r+1}}.
\enq
Since the partition function of ${\cal N}=4$ super Yang-Mills for ${\cal G}_r$ has the same modular property as $SU(|\Gamma_{\cal G}|)$,
we introduce the following formula for $ADE$ case
\beq
\frac{{\tilde \Theta}_{{\cal G}_r}^{d,b}(\tau)}{\eta(\tau)^{r+1}},
\enq
where ${\tilde \Theta}_{{\cal G}_r}^{d,b}(\tau)$ has the same modular property as
\beq
\frac{\eta(\tau)^{r+1}}{\eta(\frac{a\tau+b}{d})},(ad=|\Gamma_{\cal G}|,b<d,a,b,d\in{\bf Z}).
\enq

\subsection{$D_{2n}$}
$D_{2n}$ has center ${\bf Z}_2\times {\bf Z}_2$.
However the tensor structure of the center is not suitable for
modular property of $D_{2n}$ partition function.
As in \cite{jin3}, we consider the reduction ${\bf Z}_2\times {\bf Z}_2 \to {\bf Z}_2$.
Thus we consider theta function of modular property of $\frac{\eta(\tau)^{n+1}}{\eta(\frac{\tau}{2})}$ by using level $2~D_{2n}$ theta functions.

\paragraph{$D_2$}~\\
It is well-known that
$D_2=A_1\oplus A_1.$ On the other hand this algebra has already been considered in Sec.4.2 as a part of $A_1$.

\paragraph{$D_4$}~\\
For level $2~D_4$ theta functions, there are 4 Weyl orbits.
After some calculations, we find
\beq
 {\tilde \Theta}_{D_4}^{1,0}(\tau):=\theta_{D_4}^{\lambda_0}-\theta_{D_4}^{\lambda_1^2},
\enq
\beq
 {\tilde \Theta}_{D_4}^{2,b}(\tau):=2\zeta_{24}^{-5b}\zeta_4^{b}(\theta_{D_4}^{\lambda_1^1}+(-1)^b2\theta_{D_4}^{\lambda_3}),b=0,1.
\enq
Modular property of these theta functions are given by
\beq
{\tilde \Theta}_{D_4}^{1,0}(-\frac{1}{\tau})=(-i\tau)^22{\tilde \Theta}_{D_4}^{2,0}(\tau),
\enq
\beq
{\tilde \Theta}_{D_4}^{2,1}(-\frac{1}{\tau})=(-i\tau)^2{\tilde \Theta}_{D_4}^{2,1}(\tau).
\enq
This modular property is the same as that in (4.23) and (4.24) except for $-i\tau$ factors.
\subsection{$D_{2n+1}$}
$D_{2n+1}$ has center ${\bf Z}_4$.We consider theta function of modular property of $\frac{\eta(\tau)^{n+1}}{\eta(\frac{\tau}{4})}$ by using level $4~D_{2n+1}$ theta functions.

\paragraph{$D_3$}~\\
It is well-known that
$D_3=A_3$. Thus we have already considered $D_3$ case in Sec.4.2.

\paragraph{$D_5$}~\\
For level $4~D_5$ theta functions, there are 39 Weyl orbits.
After some trials, we find that there is identity between $\frac{\eta(\tau)^6}{\eta(\frac{\tau}{4})}$ and level $4~D_5$ theta functions like $A_3$ case. We define $\Theta_{D_5}^{d,b}(\tau)$ as
\beq
\frac{1}{\eta(\frac{a\tau+b}{d})}=:\frac{\Theta_{D_5}^{d,b}(\tau)}{\eta(\tau)^6},(ad=4,b<d,a,b,d\in{\bf Z}).
\enq
By comparing q-expansion of $(\Theta_{D_5}^{d,b}(\tau))^3$
with level $4~D_5$ theta functions, we can introduce
\beqy
&&({\tilde \Theta}_{D_5}^{1,0}(\tau))^3
\no\\&&
:=8\theta_{D_5}^{\lambda_0}\theta_{D_5}^{\lambda_2^2}\theta_{D_5}^{\lambda_{4}}
-64\theta_{D_5}^{\lambda_{5}^2}\theta_{D_5}^{\lambda_{13}}\theta_{D_5}^{\lambda_{26}}
+128\theta_{D_5}^{\lambda_{5}^2}\theta_{D_5}^{\lambda_0}\theta_{D_5}^{\lambda_{6}}
-64\theta_{D_5}^{\lambda_{15}^2}\theta_{D_5}^{\lambda_{1}^3}\theta_{D_5}^{\lambda_{9}^1}
+80\theta_{D_5}^{\lambda_{1}^2}\theta_{D_5}^{\lambda_{5}^1}\theta_{D_5}^{\lambda_{27}}
-96\theta_{D_5}^{\lambda_{1}^2}\theta_{D_5}^{\lambda_{14}^2}\theta_{D_5}^{\lambda_{1}^1}
\no\\&&
-16\theta_{D_5}^{\lambda_2^2}\theta_{D_5}^{\lambda_{28}}\theta_{D_5}^{\lambda_{14}^4}
-24\theta_{D_5}^{\lambda_1^4}\theta_{D_5}^{\lambda_2^2}\theta_{D_5}^{\lambda_{4}}
-2\theta_{D_5}^{\lambda_0}\theta_{D_5}^{\lambda_{1}^1}\theta_{D_5}^{\lambda_{1}^3}
+64\theta_{D_5}^{\lambda_{15}^2}\theta_{D_5}^{\lambda_{4}}\theta_{D_5}^{\lambda_{27}}
+192\theta_{D_5}^{\lambda_{5}^2}\theta_{D_5}^{\lambda_{14}^2}\theta_{D_5}^{\lambda_{1}^1}
-\theta_{D_5}^{\lambda_1^4}\theta_{D_5}^{\lambda_0}\theta_{D_5}^{\lambda_{2}^1}
\no\\&&
+32\theta_{D_5}^{\lambda_2^2}\theta_{D_5}^{\lambda_{26}}\theta_{D_5}^{\lambda_{1}^3}
+64\theta_{D_5}^{\lambda_{15}^2}\theta_{D_5}^{\lambda_{2}^1}\theta_{D_5}^{\lambda_{7}}
+8\theta_{D_5}^{\lambda_0}\theta_{D_5}^{\lambda_{7}}\theta_{D_5}^{\lambda_{10}}
+64\theta_{D_5}^{\lambda_{1}^2}\theta_{D_5}^{\lambda_{9}^1}\theta_{D_5}^{\lambda_{2}^1}
-24\theta_{D_5}^{\lambda_{1}^2}\theta_{D_5}^{\lambda_0}\theta_{D_5}^{\lambda_{6}}
-192\theta_{D_5}^{\lambda_{5}^2}\theta_{D_5}^{\lambda_{3}}\theta_{D_5}^{\lambda_{1}^1}
\no\\&&
+4\theta_{D_5}^{\lambda_1^4}\theta_{D_5}^{\lambda_{26}}\theta_{D_5}^{\lambda_{1}^3}
+136\theta_{D_5}^{\lambda_{1}^2}\theta_{D_5}^{\lambda_{3}}\theta_{D_5}^{\lambda_{1}^1}
+48\theta_{D_5}^{\lambda_1^4}\theta_{D_5}^{\lambda_{6}}\theta_{D_5}^{\lambda_{11}}
-16\theta_{D_5}^{\lambda_0}\theta_{D_5}^{\lambda_{6}}\theta_{D_5}^{\lambda_{11}}
-128\theta_{D_5}^{\lambda_{5}^2}\theta_{D_5}^{\lambda_{9}^1}\theta_{D_5}^{\lambda_{12}}
+4\theta_{D_5}^{\lambda_0}\theta_{D_5}^{\lambda_{26}}\theta_{D_5}^{\lambda_{1}^3}
\no\\&&
-56\theta_{D_5}^{\lambda_1^4}\theta_{D_5}^{\lambda_{7}}\theta_{D_5}^{\lambda_{10}}
+2\theta_{D_5}^{\lambda_1^4}\theta_{D_5}^{\lambda_{1}^1}\theta_{D_5}^{\lambda_{1}^3}
+192\theta_{D_5}^{\lambda_{5}^2}\theta_{D_5}^{\lambda_{15}^2}\theta_{D_5}^{\lambda_{26}}
-16\theta_{D_5}^{\lambda_{1}^2}\theta_{D_5}^{\lambda_{9}^1}\theta_{D_5}^{\lambda_{12}}
-64\theta_{D_5}^{\lambda_{15}^2}\theta_{D_5}^{\lambda_{26}}\theta_{D_5}^{\lambda_{11}}
-96\theta_{D_5}^{\lambda_{1}^2}\theta_{D_5}^{\lambda_{15}^2}\theta_{D_5}^{\lambda_{26}}
\no\\&&
+32\theta_{D_5}^{\lambda_{1}^2}\theta_{D_5}^{\lambda_{13}}\theta_{D_5}^{\lambda_{26}}
+64\theta_{D_5}^{\lambda_{15}^2}\theta_{D_5}^{\lambda_{7}}\theta_{D_5}^{\lambda_{4}}
-128\theta_{D_5}^{\lambda_{5}^2}\theta_{D_5}^{\lambda_{9}^1}\theta_{D_5}^{\lambda_{2}^1}
-64\theta_{D_5}^{\lambda_{15}^2}\theta_{D_5}^{\lambda_{5}^1}\theta_{D_5}^{\lambda_{4}}
+\theta_{D_5}^{\lambda_0}\theta_{D_5}^{\lambda_1^4}\theta_{D_5}^{\lambda_{4}}
-(\theta_{D_5}^{\lambda_1^4})^2\theta_{D_5}^{\lambda_{4}}
\no\\&&
-16\theta_{D_5}^{\lambda_2^2}(\theta_{D_5}^{\lambda_1^1})^2
+(\theta_{D_5}^{\lambda_0})^2\theta_{D_5}^{\lambda_{2}^1}
-4\theta_{D_5}^{\lambda_0}(\theta_{D_5}^{\lambda_{3}})^2
+28\theta_{D_5}^{\lambda_1^4}(\theta_{D_5}^{\lambda_{3}})^2
-60(\theta_{D_5}^{\lambda_{1}^2})^2\theta_{D_5}^{\lambda_{2}^1}
-20\theta_{D_5}^{\lambda_{4}}(\theta_{D_5}^{\lambda_{1}^2})^2,
\enqy
\beqy
&&({\tilde \Theta}_{D_5}^{2,0}(\tau))^3
\no\\&&
:=\theta_{D_5}^{\lambda_{1}^2}\theta_{D_5}^{\lambda_1^4}\theta_{D_5}^{\lambda_{1}^1}
+\theta_{D_5}^{\lambda_{1}^2}\theta_{D_5}^{\lambda_0}\theta_{D_5}^{\lambda_{1}^1}
+\theta_{D_5}^{\lambda_{1}^2}\theta_{D_5}^{\lambda_0}\theta_{D_5}^{\lambda_{1}^3}
+4\theta_{D_5}^{\lambda_{1}^2}\theta_{D_5}^{\lambda_{1}^3}\theta_{D_5}^{\lambda_2^2}
+\theta_{D_5}^{\lambda_{1}^2}\theta_{D_5}^{\lambda_1^4}\theta_{D_5}^{\lambda_{1}^3}
-48\theta_{D_5}^{\lambda_{1}^2}\theta_{D_5}^{\lambda_{28}}\theta_{D_5}^{\lambda_{27}}
\no\\&&
-48\theta_{D_5}^{\lambda_{1}^2}\theta_{D_5}^{\lambda_{28}}\theta_{D_5}^{\lambda_{5}^1}
-56\theta_{D_5}^{\lambda_{5}^2}\theta_{D_5}^{\lambda_{11}}\theta_{D_5}^{\lambda_{14}^2}
-56\theta_{D_5}^{\lambda_{5}^2}\theta_{D_5}^{\lambda_{9}^1}\theta_{D_5}^{\lambda_{14}^2}
-96\theta_{D_5}^{\lambda_{15}^2}\theta_{D_5}^{\lambda_{6}}\theta_{D_5}^{\lambda_4}
-96\theta_{D_5}^{\lambda_{15}^2}\theta_{D_5}^{\lambda_{14}^2}\theta_{D_5}^{\lambda_7}
+192\theta_{D_5}^{\lambda_{15}^2}\theta_{D_5}^{\lambda_{14}^2}\theta_{D_5}^{\lambda_{5}^1}
\no\\&&
+96\theta_{D_5}^{\lambda_{15}^2}\theta_{D_5}^{\lambda_{14}^2}\theta_{D_5}^{\lambda_{27}}
+192\theta_{D_5}^{\lambda_{15}^2}\theta_{D_5}^{\lambda_{1}^3}\theta_{D_5}^{\lambda_{10}}
+128\theta_{D_5}^{\lambda_{1}^2}\theta_{D_5}^{\lambda_{11}}\theta_{D_5}^{\lambda_{14}^2}
-144\theta_{D_5}^{\lambda_{1}^2}\theta_{D_5}^{\lambda_{6}}\theta_{D_5}^{\lambda_{5}^1}
+128\theta_{D_5}^{\lambda_{1}^2}\theta_{D_5}^{\lambda_{9}^1}\theta_{D_5}^{\lambda_{14}^2}
\no\\&&
+16\theta_{D_5}^{\lambda_{5}^2}\theta_{D_5}^{\lambda_{28}}\theta_{D_5}^{\lambda_{27}}
+80\theta_{D_5}^{\lambda_{5}^2}\theta_{D_5}^{\lambda_{6}}\theta_{D_5}^{\lambda_{5}^1}
+16\theta_{D_5}^{\lambda_{5}^2}\theta_{D_5}^{\lambda_{28}}\theta_{D_5}^{\lambda_{5}^1}
+8\theta_{D_5}^{\lambda_{14}^4}\theta_{D_5}^{\lambda_{6}}\theta_{D_5}^{\lambda_{5}^1}
-8\theta_{D_5}^{\lambda_{14}^4}\theta_{D_5}^{\lambda_{28}}\theta_{D_5}^{\lambda_{27}}
-24\theta_{D_5}^{\lambda_{1}^2}\theta_{D_5}^{\lambda_{6}}\theta_{D_5}^{\lambda_7}
\no\\&&
-96\theta_{D_5}^{\lambda_{15}^2}(\theta_{D_5}^{\lambda_{11}})^2
-192\theta_{D_5}^{\lambda_{15}^2}(\theta_{D_5}^{\lambda_{9}^1})^2,
\enqy
\beqy
&&({\tilde \Theta}_{D_5}^{2,1}(\tau))^3
\no\\&&
:=\zeta_{16}^{-1}(\theta_{D_5}^{\lambda_{1}^2}\theta_{D_5}^{\lambda_{1}^4}\theta_{D_5}^{\lambda_{1}^1}
+\theta_{D_5}^{\lambda_{1}^2}\theta_{D_5}^{\lambda_0}\theta_{D_5}^{\lambda_{1}^1}
+\theta_{D_5}^{\lambda_{1}^2}\theta_{D_5}^{\lambda_0}\theta_{D_5}^{\lambda_{1}^3}
+4\theta_{D_5}^{\lambda_{1}^2}\theta_{D_5}^{\lambda_{1}^3}\theta_{D_5}^{\lambda_{2}^2}
+\theta_{D_5}^{\lambda_{1}^2}\theta_{D_5}^{\lambda_{1}^4}\theta_{D_5}^{\lambda_{1}^3}
\no\\&&
-48\theta_{D_5}^{\lambda_{1}^2}\theta_{D_5}^{\lambda_{28}}\theta_{D_5}^{\lambda_{27}}
-48\theta_{D_5}^{\lambda_{1}^2}\theta_{D_5}^{\lambda_{28}}\theta_{D_5}^{\lambda_{5}^1}
-56\theta_{D_5}^{\lambda_{5}^2}\theta_{D_5}^{\lambda_{11}}\theta_{D_5}^{\lambda_{14}^2}
-56\theta_{D_5}^{\lambda_{5}^2}\theta_{D_5}^{\lambda_{9}^1}\theta_{D_5}^{\lambda_{14}^2}
-96\theta_{D_5}^{\lambda_{15}^2}\theta_{D_5}^{\lambda_{6}}\theta_{D_5}^{\lambda_4}
-96\theta_{D_5}^{\lambda_{15}^2}\theta_{D_5}^{\lambda_{14}^2}\theta_{D_5}^{\lambda_7}
\no\\&&
+192\theta_{D_5}^{\lambda_{15}^2}\theta_{D_5}^{\lambda_{14}^2}\theta_{D_5}^{\lambda_{5}^1}
+96\theta_{D_5}^{\lambda_{15}^2}\theta_{D_5}^{\lambda_{14}^2}\theta_{D_5}^{\lambda_{27}}
+192\theta_{D_5}^{\lambda_{15}^2}\theta_{D_5}^{\lambda_{1}^3}\theta_{D_5}^{\lambda_{10}}
+128\theta_{D_5}^{\lambda_{1}^2}\theta_{D_5}^{\lambda_{11}}\theta_{D_5}^{\lambda_{14}^2}
-144\theta_{D_5}^{\lambda_{1}^2}\theta_{D_5}^{\lambda_{6}}\theta_{D_5}^{\lambda_{5}^1}
\no\\&&
+128\theta_{D_5}^{\lambda_{1}^2}\theta_{D_5}^{\lambda_{9}^1}\theta_{D_5}^{\lambda_{14}^2}
+16\theta_{D_5}^{\lambda_{5}^2}\theta_{D_5}^{\lambda_{28}}\theta_{D_5}^{\lambda_{27}}
+80\theta_{D_5}^{\lambda_{5}^2}\theta_{D_5}^{\lambda_{6}}\theta_{D_5}^{\lambda_{5}^1}
+16\theta_{D_5}^{\lambda_{5}^2}\theta_{D_5}^{\lambda_{28}}\theta_{D_5}^{\lambda_{5}^1}
+8\theta_{D_5}^{\lambda_{14}^4}\theta_{D_5}^{\lambda_{6}}\theta_{D_5}^{\lambda_{5}^1}
-8\theta_{D_5}^{\lambda_{14}^4}\theta_{D_5}^{\lambda_{28}}\theta_{D_5}^{\lambda_{27}}
\no\\&&
-24\theta_{D_5}^{\lambda_{1}^2}\theta_{D_5}^{\lambda_{6}}\theta_{D_5}^{\lambda_7}
-96\theta_{D_5}^{\lambda_{15}^2}(\theta_{D_5}^{\lambda_{11}})^2
-192\theta_{D_5}^{\lambda_{15}^2}(\theta_{D_5}^{\lambda_{9}^1})^2),
\enqy
\beqy
&&({\tilde \Theta}_{D_5}^{4,b}(\tau))^3
\no\\&&
:=\zeta_4^{-3b}\zeta_{32}^{23b}((1729(\theta_{D_5}^{\lambda_{16}^3})^2\theta_{D_5}^{\lambda_{14}^3}
-215(\theta_{D_5}^{\lambda_{20}})^2\theta_{D_5}^{\lambda_{23}}
+(\theta_{D_5}^{\lambda_{14}^1})^2\theta_{D_5}^{\lambda_{15}^1}
-904\theta_{D_5}^{\lambda_{20}}\theta_{D_5}^{\lambda_{23}}\theta_{D_5}^{\lambda_{16}^3}
-1265(\theta_{D_5}^{\lambda_{21}})^2\theta_{D_5}^{\lambda_{23}}
\no\\&&
+3(\theta_{D_5}^{\lambda_{17}})^2\theta_{D_5}^{\lambda_{15}^1}
+3\theta_{D_5}^{\lambda_{22}}(\theta_{D_5}^{\lambda_{14}^3})^2
+33\theta_{D_5}^{\lambda_{22}}(\theta_{D_5}^{\lambda_{15}^1})^2
+1149\theta_{D_5}^{\lambda_{22}}\theta_{D_5}^{\lambda_{21}}\theta_{D_5}^{\lambda_{20}}
-297\theta_{D_5}^{\lambda_{22}}\theta_{D_5}^{\lambda_{15}^1}\theta_{D_5}^{\lambda_{14}^3}
\no\\&&
-1742\theta_{D_5}^{\lambda_{18}}\theta_{D_5}^{\lambda_{23}}\theta_{D_5}^{\lambda_{14}^3}
+119\theta_{D_5}^{\lambda_{17}}\theta_{D_5}^{\lambda_{21}}\theta_{D_5}^{\lambda_{23}}
-82\theta_{D_5}^{\lambda_{19}}\theta_{D_5}^{\lambda_{17}}\theta_{D_5}^{\lambda_{23}}
+26\theta_{D_5}^{\lambda_{19}}\theta_{D_5}^{\lambda_{17}}\theta_{D_5}^{\lambda_{15}^1}
+22\theta_{D_5}^{\lambda_{18}}\theta_{D_5}^{\lambda_{21}}\theta_{D_5}^{\lambda_{16}^3}
\no\\&&
+1329\theta_{D_5}^{\lambda_{18}}\theta_{D_5}^{\lambda_{21}}\theta_{D_5}^{\lambda_{20}}
+85\theta_{D_5}^{\lambda_{22}}(\theta_{D_5}^{\lambda_{16}^1})^2
+6(\theta_{D_5}^{\lambda_{16}^1})^3)
\no\\&&
+\zeta_4^b(9(\theta_{D_5}^{\lambda_{21}})^3
+(\theta_{D_5}^{\lambda_{18}})^2\theta_{D_5}^{\lambda_{16}^3}
+3\theta_{D_5}^{\lambda_{21}}\theta_{D_5}^{\lambda_{18}}\theta_{D_5}^{\lambda_{14}^3}
+321\theta_{D_5}^{\lambda_{19}}\theta_{D_5}^{\lambda_{22}}\theta_{D_5}^{\lambda_{14}^3}
+747\theta_{D_5}^{\lambda_{22}}\theta_{D_5}^{\lambda_{18}}\theta_{D_5}^{\lambda_{20}}
\no\\&&
+60\theta_{D_5}^{\lambda_{16}^1}\theta_{D_5}^{\lambda_{22}}\theta_{D_5}^{\lambda_{14}^1}
-428\theta_{D_5}^{\lambda_{16}^1}\theta_{D_5}^{\lambda_{22}}\theta_{D_5}^{\lambda_{16}^3}
+1329\theta_{D_5}^{\lambda_{18}}\theta_{D_5}^{\lambda_{21}}\theta_{D_5}^{\lambda_{20}}
+26\theta_{D_5}^{\lambda_{19}}\theta_{D_5}^{\lambda_{17}}\theta_{D_5}^{\lambda_{15}^1}
-82\theta_{D_5}^{\lambda_{19}}\theta_{D_5}^{\lambda_{17}}\theta_{D_5}^{\lambda_{23}}
\no\\&&
+119\theta_{D_5}^{\lambda_{17}}\theta_{D_5}^{\lambda_{21}}\theta_{D_5}^{\lambda_{23}}
-1742\theta_{D_5}^{\lambda_{18}}\theta_{D_5}^{\lambda_{23}}\theta_{D_5}^{\lambda_{14}^3}
-297\theta_{D_5}^{\lambda_{22}}\theta_{D_5}^{\lambda_{15}^1}\theta_{D_5}^{\lambda_{14}^3}
+1149\theta_{D_5}^{\lambda_{22}}\theta_{D_5}^{\lambda_{21}}\theta_{D_5}^{\lambda_{20}}
-904\theta_{D_5}^{\lambda_{20}}\theta_{D_5}^{\lambda_{23}}\theta_{D_5}^{\lambda_{16}^3}
\no\\&&
-9\theta_{D_5}^{\lambda_{16}^3}(\theta_{D_5}^{\lambda_{15}^1})^2
+70\theta_{D_5}^{\lambda_{20}}(\theta_{D_5}^{\lambda_{23}})^2
+3\theta_{D_5}^{\lambda_{19}}(\theta_{D_5}^{\lambda_{14}^1})^2
+100\theta_{D_5}^{\lambda_{22}}\theta_{D_5}^{\lambda_{18}}\theta_{D_5}^{\lambda_{14}^1}
+22\theta_{D_5}^{\lambda_{18}}\theta_{D_5}^{\lambda_{21}}\theta_{D_5}^{\lambda_{16}^3}
\no\\&&
+1729(\theta_{D_5}^{\lambda_{16}^3})^2\theta_{D_5}^{\lambda_{14}^3}
-215(\theta_{D_5}^{\lambda_{20}})^2\theta_{D_5}^{\lambda_{23}}
-1265(\theta_{D_5}^{\lambda_{21}})^2\theta_{D_5}^{\lambda_{23}}
+3(\theta_{D_5}^{\lambda_{17}})^2\theta_{D_5}^{\lambda_{15}^1}
+3\theta_{D_5}^{\lambda_{22}}(\theta_{D_5}^{\lambda_{14}^3})^2
\no\\&&
+4\theta_{D_5}^{\lambda_{17}}(\theta_{D_5}^{\lambda_{20}})^2
+10\theta_{D_5}^{\lambda_{16}^3}(\theta_{D_5}^{\lambda_{23}})^2
+23\theta_{D_5}^{\lambda_{21}}(\theta_{D_5}^{\lambda_{16}^3})^2
+33\theta_{D_5}^{\lambda_{22}}(\theta_{D_5}^{\lambda_{15}^1})^2
+(\theta_{D_5}^{\lambda_{14}^1})^2\theta_{D_5}^{\lambda_{15}^1}
\no\\&&
-1042(\theta_{D_5}^{\lambda_{22}})^2\theta_{D_5}^{\lambda_{16}^3}
+100(\theta_{D_5}^{\lambda_{21}})^2\theta_{D_5}^{\lambda_{17}}
-43(\theta_{D_5}^{\lambda_{16}^1})^2\theta_{D_5}^{\lambda_{16}^3}
+9(\theta_{D_5}^{\lambda_{17}})^2\theta_{D_5}^{\lambda_{19}}
+62(\theta_{D_5}^{\lambda_{19}})^2\theta_{D_5}^{\lambda_{21}})
\no\\&&
+\zeta_4^{2b}(\theta_{D_5}^{\lambda_{16}^1}(\theta_{D_5}^{\lambda_{17}})^2
-41\theta_{D_5}^{\lambda_{17}}\theta_{D_5}^{\lambda_{18}}\theta_{D_5}^{\lambda_{21}}
+33\theta_{D_5}^{\lambda_{21}}\theta_{D_5}^{\lambda_{17}}\theta_{D_5}^{\lambda_{16}^1}
-50\theta_{D_5}^{\lambda_{21}}\theta_{D_5}^{\lambda_{20}}\theta_{D_5}^{\lambda_{23}}
-48\theta_{D_5}^{\lambda_{21}}\theta_{D_5}^{\lambda_{16}^3}\theta_{D_5}^{\lambda_{23}}
\no\\&&
-51\theta_{D_5}^{\lambda_{21}}\theta_{D_5}^{\lambda_{16}^3}\theta_{D_5}^{\lambda_{14}^3}
+30\theta_{D_5}^{\lambda_{22}}\theta_{D_5}^{\lambda_{14}^1}\theta_{D_5}^{\lambda_{16}^3}
+3\theta_{D_5}^{\lambda_{22}}\theta_{D_5}^{\lambda_{14}^1}\theta_{D_5}^{\lambda_{20}}
+23\theta_{D_5}^{\lambda_{17}}\theta_{D_5}^{\lambda_{14}^1}\theta_{D_5}^{\lambda_{23}}
-107\theta_{D_5}^{\lambda_{19}}\theta_{D_5}^{\lambda_{20}}\theta_{D_5}^{\lambda_{23}}
\no\\&&
+3\theta_{D_5}^{\lambda_{17}}\theta_{D_5}^{\lambda_{14}^1}\theta_{D_5}^{\lambda_{15}^1}
-6\theta_{D_5}^{\lambda_{17}}\theta_{D_5}^{\lambda_{16}^3}\theta_{D_5}^{\lambda_{14}^3}
+19\theta_{D_5}^{\lambda_{18}}\theta_{D_5}^{\lambda_{20}}\theta_{D_5}^{\lambda_{16}^3}
+1329\theta_{D_5}^{\lambda_{18}}\theta_{D_5}^{\lambda_{21}}\theta_{D_5}^{\lambda_{20}}
+26\theta_{D_5}^{\lambda_{19}}\theta_{D_5}^{\lambda_{17}}\theta_{D_5}^{\lambda_{15}^1}
\no\\&&
-82\theta_{D_5}^{\lambda_{19}}\theta_{D_5}^{\lambda_{17}}\theta_{D_5}^{\lambda_{23}}
+119\theta_{D_5}^{\lambda_{17}}\theta_{D_5}^{\lambda_{21}}\theta_{D_5}^{\lambda_{23}}
-1742\theta_{D_5}^{\lambda_{18}}\theta_{D_5}^{\lambda_{23}}\theta_{D_5}^{\lambda_{14}^3}
-297\theta_{D_5}^{\lambda_{22}}\theta_{D_5}^{\lambda_{15}^1}\theta_{D_5}^{\lambda_{14}^3}
+1149\theta_{D_5}^{\lambda_{22}}\theta_{D_5}^{\lambda_{21}}\theta_{D_5}^{\lambda_{20}}
\no\\&&
-904\theta_{D_5}^{\lambda_{20}}\theta_{D_5}^{\lambda_{23}}\theta_{D_5}^{\lambda_{16}^3}
+22\theta_{D_5}^{\lambda_{18}}\theta_{D_5}^{\lambda_{21}}\theta_{D_5}^{\lambda_{16}^3}
+9\theta_{D_5}^{\lambda_{22}}(\theta_{D_5}^{\lambda_{14}^1})^2
+1729(\theta_{D_5}^{\lambda_{16}^3})^2\theta_{D_5}^{\lambda_{14}^3}
-215(\theta_{D_5}^{\lambda_{20}})^2\theta_{D_5}^{\lambda_{23}}
\no\\&&
-1265(\theta_{D_5}^{\lambda_{21}})^2\theta_{D_5}^{\lambda_{23}}
+3(\theta_{D_5}^{\lambda_{17}})^2\theta_{D_5}^{\lambda_{15}^1}
+3\theta_{D_5}^{\lambda_{22}}(\theta_{D_5}^{\lambda_{14}^3})^2
+33\theta_{D_5}^{\lambda_{22}}(\theta_{D_5}^{\lambda_{15}^1})^2
+(\theta_{D_5}^{\lambda_{14}^1})^2\theta_{D_5}^{\lambda_{15}^1}
\no\\&&
+140(\theta_{D_5}^{\lambda_{23}})^2\theta_{D_5}^{\lambda_{14}^3}
+60\theta_{D_5}^{\lambda_{23}}(\theta_{D_5}^{\lambda_{14}^3})^2
+45\theta_{D_5}^{\lambda_{16}^1}(\theta_{D_5}^{\lambda_{16}^3})^2
-47\theta_{D_5}^{\lambda_{15}^1}(\theta_{D_5}^{\lambda_{14}^3})^2
+3\theta_{D_5}^{\lambda_{18}}(\theta_{D_5}^{\lambda_{16}^3})^2
-19(\theta_{D_5}^{\lambda_{15}^1})^2\theta_{D_5}^{\lambda_{14}^3})
\no\\&&
+\zeta_4^{3b}(\theta_{D_5}^{\lambda_{21}}(\theta_{D_5}^{\lambda_{14}^3})^2
-28\theta_{D_5}^{\lambda_{21}}\theta_{D_5}^{\lambda_{19}}\theta_{D_5}^{\lambda_{16}^3}
-18\theta_{D_5}^{\lambda_{19}}\theta_{D_5}^{\lambda_{15}^1}\theta_{D_5}^{\lambda_{14}^3}
-17\theta_{D_5}^{\lambda_{18}}\theta_{D_5}^{\lambda_{20}}\theta_{D_5}^{\lambda_{14}^3}
+61\theta_{D_5}^{\lambda_{18}}\theta_{D_5}^{\lambda_{20}}\theta_{D_5}^{\lambda_{23}}
\no\\&&
-7\theta_{D_5}^{\lambda_{18}}\theta_{D_5}^{\lambda_{14}^1}\theta_{D_5}^{\lambda_{14}^3}
+13\theta_{D_5}^{\lambda_{22}}\theta_{D_5}^{\lambda_{14}^1}\theta_{D_5}^{\lambda_{15}^1}
+160\theta_{D_5}^{\lambda_{22}}\theta_{D_5}^{\lambda_{16}^3}\theta_{D_5}^{\lambda_{14}^3}
-75\theta_{D_5}^{\lambda_{22}}\theta_{D_5}^{\lambda_{16}^3}\theta_{D_5}^{\lambda_{15}^1}
+3\theta_{D_5}^{\lambda_{18}}\theta_{D_5}^{\lambda_{16}^3}\theta_{D_5}^{\lambda_{14}^3}
\no\\&&
+9\theta_{D_5}^{\lambda_{18}}\theta_{D_5}^{\lambda_{16}^3}\theta_{D_5}^{\lambda_{23}}
+4\theta_{D_5}^{\lambda_{16}^1}\theta_{D_5}^{\lambda_{14}^1}\theta_{D_5}^{\lambda_{14}^3}
+9\theta_{D_5}^{\lambda_{16}^1}\theta_{D_5}^{\lambda_{14}^1}\theta_{D_5}^{\lambda_{15}^1}
+17\theta_{D_5}^{\lambda_{16}^1}\theta_{D_5}^{\lambda_{16}^3}\theta_{D_5}^{\lambda_{14}^3}
-3\theta_{D_5}^{\lambda_{16}^1}\theta_{D_5}^{\lambda_{20}}\theta_{D_5}^{\lambda_{14}^3}
\no\\&&
+1329\theta_{D_5}^{\lambda_{18}}\theta_{D_5}^{\lambda_{21}}\theta_{D_5}^{\lambda_{20}}
+26\theta_{D_5}^{\lambda_{19}}\theta_{D_5}^{\lambda_{17}}\theta_{D_5}^{\lambda_{15}^1}
-82\theta_{D_5}^{\lambda_{19}}\theta_{D_5}^{\lambda_{17}}\theta_{D_5}^{\lambda_{23}}
+119\theta_{D_5}^{\lambda_{17}}\theta_{D_5}^{\lambda_{21}}\theta_{D_5}^{\lambda_{23}}
-1742\theta_{D_5}^{\lambda_{18}}\theta_{D_5}^{\lambda_{23}}\theta_{D_5}^{\lambda_{14}^3}
\no\\&&
-297\theta_{D_5}^{\lambda_{22}}\theta_{D_5}^{\lambda_{15}^1}\theta_{D_5}^{\lambda_{14}^3}
+1149\theta_{D_5}^{\lambda_{22}}\theta_{D_5}^{\lambda_{21}}\theta_{D_5}^{\lambda_{20}}
-904\theta_{D_5}^{\lambda_{20}}\theta_{D_5}^{\lambda_{23}}\theta_{D_5}^{\lambda_{16}^3}
+22\theta_{D_5}^{\lambda_{18}}\theta_{D_5}^{\lambda_{21}}\theta_{D_5}^{\lambda_{16}^3}
-7\theta_{D_5}^{\lambda_{17}}(\theta_{D_5}^{\lambda_{14}^3})^2
\no\\&&
-56\theta_{D_5}^{\lambda_{17}}(\theta_{D_5}^{\lambda_{23}})^2
+1729(\theta_{D_5}^{\lambda_{16}^3})^2\theta_{D_5}^{\lambda_{14}^3}
-215(\theta_{D_5}^{\lambda_{20}})^2\theta_{D_5}^{\lambda_{23}}
-1265(\theta_{D_5}^{\lambda_{21}})^2\theta_{D_5}^{\lambda_{23}}
+3(\theta_{D_5}^{\lambda_{17}})^2\theta_{D_5}^{\lambda_{15}^1}
\no\\&&
+3\theta_{D_5}^{\lambda_{22}}(\theta_{D_5}^{\lambda_{14}^3})^2
+33\theta_{D_5}^{\lambda_{22}}(\theta_{D_5}^{\lambda_{15}^1})^2
+(\theta_{D_5}^{\lambda_{14}^1})^2\theta_{D_5}^{\lambda_{15}^1}
+3\theta_{D_5}^{\lambda_{17}}(\theta_{D_5}^{\lambda_{16}^1})^2
+47\theta_{D_5}^{\lambda_{19}}(\theta_{D_5}^{\lambda_{16}^1})^2
\no\\&&
+4(\theta_{D_5}^{\lambda_{17}})^2\theta_{D_5}^{\lambda_{16}^3}
-120(\theta_{D_5}^{\lambda_{21}})^2\theta_{D_5}^{\lambda_{16}^3})),b=0,1,2,3,
\enqy
where q-expansion of $({\tilde \Theta}_{D_5}^{d,b}(\tau))^3$ is the same as that of $(\Theta_{D_5}^{d,b}(\tau))^3$.
In the same way as $A_3$, the expression (4.25)...(4.28) has an ambiguity because of the identities between theta functions.
On the other hands we can also investigate modular property of these theta functions by using (4.6). We obtain
\beq
({\tilde \Theta}_{D_5}^{1,0}(-\frac{1}{\tau}))^3=(-i\tau)^{\frac{15}{2}}2^3({\tilde \Theta}_{D_5}^{4,0}(\tau))^3,
\enq
\beq
({\tilde \Theta}_{D_5}^{2,0}(-\frac{1}{\tau}))^3=(-i\tau)^{\frac{15}{2}}({\tilde \Theta}_{D_5}^{2,0}(\tau))^3,
\enq
\beq
({\tilde \Theta}_{D_5}^{2,1}(-\frac{1}{\tau}))^3=(-i\tau)^{\frac{15}{2}}2\sqrt{2}({\tilde \Theta}_{D_5}^{4,2}(\tau))^3,
\enq
\beq
({\tilde \Theta}_{D_5}^{4,1}(-\frac{1}{\tau}))^3=(-i\tau)^{\frac{15}{2}}\zeta_{4}({\tilde \Theta}_{D_5}^{4,3}(\tau))^3.
\enq
This modular property is the same as that of $(\Theta_{D_5}^{d,b}(\tau))^3$.
Thus we conclude the equivalence between $(\Theta_{D_5}^{d,b}(\tau))^3$ and $({\tilde\Theta}_{D_5}^{d,b}(\tau))^3$ completely.
\subsection{$E_n$}
$E_6,E_7$ and $E_8$ have center ${\bf Z}_3,{\bf Z}_2$ and ${\bf Z}_1$, respectively, and the corresponding $SU(N)$ theories are $SU(3),SU(2)$ and $U(1)$. We search for theta functions of the same modular property as $\frac{\eta(\tau)^{r+1}}{\eta(\frac{\tau}{|\Gamma_{E_r}|})}$ by using level $|\Gamma_{E_r}|~E_r$ theta functions.
\paragraph{$E_6$}~\\
For level $3~E_6$ theta functions, there are 11 Weyl orbits.
By imitating $A_2$ case, we introduce
\beqy
{\tilde \Theta}_{E_6}^{1,0}(\tau)&:=&
\theta_{E_6}^{\lambda_0}-\theta_{E_6}^{\lambda_1^3},
\enqy
\beqy
{\tilde \Theta}_{E_6}^{3,b}(\tau)&:=&
\zeta_{24}^{-7b}\zeta_9^{2b}((\theta_{E_6}^{\lambda_1^1}+8\theta_{E_6}^{\lambda_2^2})
+\zeta_3^b(\theta_{E_6}^{\lambda_1^4}+8\theta_{E_6}^{\lambda_2^1})
+\zeta_3^{2b}(\theta_{E_6}^{\lambda_1^2}+8\theta_{E_6}^{\lambda_2^4})), \no
\\
&&b=0,1,2.
\enqy
Modular property of this theta function is given by
\beq
{\tilde \Theta}_{E_6}^{1,0}(-\frac{1}{\tau})=(-i\tau)^3\sqrt{3}{\tilde \Theta}_{E_6}^{3,0}({\tau}),
\enq
\beq
{\tilde \Theta}_{E_6}^{3,1}(-\frac{1}{\tau})=(-i\tau)^3\zeta_{24}{\tilde \Theta}_{E_6}^{3,2}({\tau}).
\enq
One find that this modular property is the same as that of (4.12) and (4.13) except for $-i\tau$ factors.

\paragraph{$E_7$}~\\
For level $2~E_7$ theta functions, there are 6 Weyl orbits.
In the same way as $A_1$ case, we consider $({\tilde \Theta}_{E_7}^{d,b}(\tau))^2$.

\beqy
({\tilde \Theta}_{E_7}^{1,0}(\tau))^2&:=&
(\theta_{E_7}^{\lambda_0})^2-(\theta_{E_7}^{\lambda_3^2})^2
-(\theta_{E_7}^{\lambda_1})^2+(\theta_{E_7}^{\lambda_2})^2
-2\theta_{E_7}^{\lambda_0}\theta_{E_7}^{\lambda_2}
+2\theta_{E_7}^{\lambda_1}\theta_{E_7}^{\lambda_3^2},
\enqy
\beqy
({\tilde \Theta}_{E_7}^{2,b}(\tau))^2&:=&
32\zeta_3^{-2b}\zeta_8^{7b}((\theta_{E_7}^{\lambda_1}\theta_{E_7}^{\lambda_3^1}+\theta_{E_7}^{\lambda_2}\theta_{E_7}^{\lambda_4})
+(-1)^b(\theta_{E_7}^{\lambda_1}\theta_{E_7}^{\lambda_4}+\theta_{E_7}^{\lambda_2}\theta_{E_7}^{\lambda_3^1})).
\enqy
Modular property of these theta functions are given by
\beq
({\tilde \Theta}_{E_7}^{1,0}(-\frac{1}{\tau}))^2=(-i\tau)^{\frac{7}{2}}2({\tilde \Theta}_{E_7}^{2,0}({\tau}))^2,
\enq
\beq
({\tilde \Theta}_{E_7}^{2,1}(-\frac{1}{\tau}))^2=(-i\tau)^{\frac{7}{2}}({\tilde \Theta}_{E_7}^{2,1}({\tau}))^2.
\enq
We also find another set of theta functions
\beqy
&&({\tilde{\tilde \Theta}}_{E_7}^{1,0}(\tau))^2
\no\\&&
:=(\theta_{E_7}^{\lambda_0})^2-(\theta_{E_7}^{\lambda_3^2})^2
-3969(\theta_{E_7}^{\lambda_1})^2+3969(\theta_{E_7}^{\lambda_2})^2
+126\theta_{E_7}^{\lambda_0}\theta_{E_7}^{\lambda_2}
-126\theta_{E_7}^{\lambda_1}\theta_{E_7}^{\lambda_3^2},
\enqy
\beqy
({\tilde{\tilde \Theta}}_{E_7}^{2,b}(\tau))^2&:=&
32\zeta_3^{-2b}\zeta_8^{7b}((7\theta_{E_7}^{\lambda_0}\theta_{E_7}^{\lambda_3^1}-9\theta_{E_7}^{\lambda_3^2}\theta_{E_7}^{\lambda_4})
+(-1)^b(7\theta_{E_7}^{\lambda_3^1}\theta_{E_7}^{\lambda_3^2}-9\theta_{E_7}^{\lambda_0}\theta_{E_7}^{\lambda_4})).
\enqy
Modular property of these theta function is also given by
\beq
({\tilde{\tilde \Theta}}_{E_7}^{1,0}(-\frac{1}{\tau}))^2=(-i\tau)^{\frac{7}{2}}2({\tilde{\tilde \Theta}}_{E_7}^{2,0}({\tau}))^2,
\enq
\beq
({\tilde{\tilde \Theta}}_{E_7}^{2,1}(-\frac{1}{\tau}))^2=(-i\tau)^{\frac{7}{2}}({\tilde{\tilde \Theta}}_{E_7}^{2,1}({\tau}))^2.
\enq
We find two sets of theta functions, which transform in the same way as $(\Theta_{A_1}^{d,b}(\tau))^2$.
To find these sets we find self-dual theta function $({\tilde\Theta}^{2,1}_{E_7}(\tau))^2~(({\tilde{\tilde\Theta}}^{2,1}_{E_7}(\tau))^2)$ at first.
Next we obtain $({\tilde\Theta}^{2,0}_{E_7}(\tau))^2~(({\tilde{\tilde\Theta}}^{2,0}_{E_7}(\tau))^2)$ by translating this.
Finally we obtain $({\tilde\Theta}^{1,0}_{E_7}(\tau))^2~(({\tilde{\tilde\Theta}}^{1,0}_{E_7}(\tau))^2)$ by modular transformation of $({\tilde\Theta}^{2,0}_{E_7}(\tau))^2~(({\tilde{\tilde\Theta}}^{2,0}_{E_7}(\tau))^2)$.

\paragraph{$E_8$}~\\
For level one $E_8$ theta function, there is only one orbit,
\beqy
{}^t\lambda_0&=&(0,0,0,0,0,0,0,0).
\enqy
We consider
\beqy
{\tilde \Theta}_{E_8}^{1,0}(\tau)&:=&
\theta_{E_8}^{\lambda_0}.
\enqy
This transforms as
\beq
{\tilde \Theta}_{E_8}^{1,0}(-\frac{1}{\tau})=(-i\tau)^4{\tilde \Theta}_{E_8}^{1,0}({\tau}).
\enq

In this section we provided $D,E$ theta function ${\tilde \Theta_{{\cal G}_r}^{d,b}}(\tau)$ which transform in the same way as $\eta(\tau)^{r+1}/\eta(\frac{a\tau+b}{d}),(a,b,d\in{\bf Z},b<d,ad=|\Gamma_{\cal G}|)$. By using this theta function we can introduce the formula for $D,E$ case ${\tilde \Theta_{{\cal G}_r}^{d,b}}(\tau)/\eta(\tau)^{r+1}$.
By considering 24-th power of this formula and combining this in the same way as the $SU(|\Gamma_{\cal G}|)$ partition function,
we obtain $D,E$ partition function which satisfies Montonen-Olive duality.
However the other Vafa-Witten conjectures are not satisfied by this approach in general.

\section{Conclusion and Discussion}
We claim that there exist at least a set of ${\cal G}_r$ theta functions $\{ {\tilde \Theta}_{{\cal G}_r}^{d,b}(\tau)|b<d \}$ constructed
by level $|\Gamma_{{\cal G}_r}|~{\cal G}_r$ theta functions, whose modular property is the same as
$\{\eta(\tau)^{r+1}/\eta(\frac{a\tau+b}{d})| ad=|\Gamma_{{\cal G}_r}|,b<d,a,b,d\in {\bf Z} \}$.
In particular for ${\cal G}_r=A_{N-1}$ case each sets are exactly the same. By considering 24-th power of ${\tilde \Theta}_{{\cal G}_r}^{d,b}(\tau)/\eta(\tau)^{r+1}$, we can construct ${\cal G}_r$ partition functions satisfying Montonen-Olive duality.
However this partition function does not satisfy the other Vafa-Witten conjectures in general.

To satisfy the other Vafa-Witten conjectures, we should search for another set of ${\cal G}_r$ functions, whose modular property is the same as
$\{\eta(\tau)^{r+1}/\eta(\frac{a\tau+b}{d})| ad=|\Gamma_{{\cal G}_r}|,b<d,a,b,d\in {\bf Z} \}$. We also attempt to search for ${\cal G}_r$ theta functions satisfying Montonen-Olive duality by using higher level theta functions.

Since we introduce a formula for $ADE$ case described by level $|\Gamma_{\cal G}|$ theta functions and produce the candidate of the $ADE$ partition function on $K3$, we want to interpret the $ADE$ partition function on $K3$ in terms of the character of affine Lie algebra in the same way as $A_1$ case.

{\bf Acknowledgment}\\
We would like to thank Prof. H.Awata, Prof. M.Jinzenji, Prof. H.Kanno, Prof. A.Tsuchiya,	Prof. M.Wakimoto and Prof. K.Yoshioka for helpful suggestions and useful discussions.
 We also thank Prof. H.Awata for carefully reading our manuscript.

\appendix
\renewcommand{\theequation}{\alph{section}.\arabic{equation}}
\section{Theta Functions}
\setcounter{equation}{0}
In this section, we give explicit notation of theta functions
 used in this article.
 For level $l$ and rank $r~{\cal G}_r$ theta function with weight vector $\beta$, we use the following notation \cite{kac, wakimoto1}:
  \beq
\theta_{{\cal G}_r}^{\beta}(\tau):=\sum_{m\in{\bf Z}^{r}}q^{\frac{l}{2}{}^t(m+\frac{1}{l}{\cal G}_r^{-1}\beta){\cal G}_r(m+\frac{1}{l}{\cal G}_r^{-1}\beta)},{}^t\beta=(b_1,b_2,\ldots,b_{r}),b_i\in {\bf Z}.
\enq
Here ${\cal G}_r$ also stands for Cartan matrix for ${\cal G}_r$.
We give independent orbits of theta functions.
\paragraph{Level $3~A_2$ theta functions}~\\
There are 6 Weyl orbits,
\beqy
{}^t\lambda_0&=&(0,0),\no\\
{}^t\lambda_1^k&=&k(1,0),k=1,2,3,4,\no\\
{}^t\lambda_2&=&(1,1).
\enqy

\paragraph{Level $4~A_3$ theta functions}~\\
There are 22 Weyl orbits,

\beqy
{}^t\lambda_0&=&(0,0,0),\no\\
{}^t\lambda_1^k&=&k(1,0,0),k=1,2,\ldots,8,\no\\
{}^t\lambda_2^k&=&k(0,1,0),k=1,2,3,\no\\
{}^t\lambda_3^k&=&k(1,1,0),k=1,3,5,7,\no\\
{}^t\lambda_4^k&=&k(1,0,1),k=1,2,\no\\
{}^t\lambda_5&=&(1,1,1),\no\\
{}^t\lambda_6&=&(2,1,0),\no\\
{}^t\lambda_7&=&(4,1,0),\no\\
{}^t\lambda_8&=&(1,4,1).
\enqy

\paragraph{Level $5~A_4$ theta functions}~\\
There are 66 Weyl orbits,
\beqy
{}^t\lambda_0&=&(0,0,0,0),\no\\
{}^t\lambda_1^k&=&k(1,0,0,0),k=1,\ldots,4,6,\ldots,9,11,12,\no\\
{}^t\lambda_2^k&=&k(0,1,0,0),k=1,\ldots,4,6,\ldots,9,11,12,\no\\
{}^t\lambda_3^k&=&k(1,1,0,0),k=1,\ldots,4,6,\ldots,9,11,12,\no\\
{}^t\lambda_4^k&=&k(1,0,1,0),k=1,\ldots,4,6,\ldots,9,11,12,\no\\
{}^t\lambda_5^k&=&k(1,1,1,0),k=1,\ldots,4,6,\ldots,9,11,12,\no\\
{}^t\lambda_1^k&=&k(1,0,0,0),k=5,10,\no\\
{}^t\lambda_6^k&=&k(1,0,0,1),k=1,2,\no\\
{}^t\lambda_7^k&=&k(2,0,1,0),k=1,2,\no\\
{}^t\lambda_8^k&=&k(0,1,1,0),k=1,2,\no\\
{}^t\lambda_9&=&(1,1,1,1),\no\\
{}^t\lambda_{10}^k&=&k(3,1,0,0),k=1,2,\no\\
{}^t\lambda_{11}^k&=&k(1,2,0,0),k=1,2,\no\\
{}^t\lambda_{12}^k&=&k(1,0,3,0),k=1,2.
\enqy.

\paragraph{Level $2~D_4$ theta functions}~\\
There are 5 Weyl orbits,
\beqy
{}^t\lambda_0&=&(0,0,0,0),\no\\
{}^t\lambda_1^k&=&k(1,0,0,0),k=1,2,\no\\
{}^t\lambda_2&=&(0,1,0,0),\no\\
{}^t\lambda_3&=&(1,0,1,0).
\enqy

\paragraph{Level $4~D_5$ theta functions}~\\
There are 39 Weyl orbits,
\beqy
{}^t\lambda_0&=&(0,0,0,0,0),\no\\
{}^t\lambda_1^k&=&k(1,0,0,0,0),k=1,2,3,4,\no\\
{}^t\lambda_2^k&=&k(0,1,0,0,0),k=1,2,\no\\
{}^t\lambda_3&=&(1,1,0,0,0),\no\\
{}^t\lambda_4&=&(2,1,0,0,0),\no\\
{}^t\lambda_5^k&=&k(0,0,1,0,0),k=1,2,\no\\
{}^t\lambda_6&=&(1,0,1,0,0),\no\\
{}^t\lambda_7&=&(2,0,1,0,0),\no\\
{}^t\lambda_8&=&(0,1,1,0,0),\no\\
{}^t\lambda_9^k&=&k(0,0,0,1,1),k=1,2,\no\\
{}^t\lambda_{10}&=&(1,0,0,1,1),\no\\
{}^t\lambda_{11}&=&(2,0,0,1,1),\no\\
{}^t\lambda_{12}&=&(0,1,0,1,1),\no\\
{}^t\lambda_{13}&=&(0,0,1,1,1),\no\\
{}^t\lambda_{14}^k&=&k(0,0,0,1,0),k=1,2,3,4,\no\\
{}^t\lambda_{15}^k&=&k(1,0,0,1,0),k=1,2,\no\\
{}^t\lambda_{16}^k&=&k(2,0,0,1,0),k=1,3,\no\\
{}^t\lambda_{17}&=&(3,0,0,1,0),\no\\
{}^t\lambda_{18}&=&(1,0,0,3,0),\no\\
{}^t\lambda_{19}&=&(0,1,0,1,0),\no\\
{}^t\lambda_{20}&=&(1,1,0,1,0),\no\\
{}^t\lambda_{21}&=&(3,2,0,3,0),\no\\
{}^t\lambda_{22}&=&(0,0,1,1,0),\no\\
{}^t\lambda_{23}&=&(1,0,1,1,0),\no\\
{}^t\lambda_{24}&=&(1,1,0,0,0),\no\\
{}^t\lambda_{25}&=&(1,0,0,2,0),\no\\
{}^t\lambda_{26}&=&(1,0,0,4,0),\no\\
{}^t\lambda_{27}&=&(0,1,0,2,0),\no\\
{}^t\lambda_{28}&=&(0,0,1,2,0).
\enqy

\paragraph{Level $3~E_6$ theta functions}~\\
There are 11 Weyl orbits,
\beqy
{}^t\lambda_0&=&(0,0,0,0,0,0),\no\\
{}^t\lambda_1^k&=&k(1,0,0,0,0,0),k=1,2,3,4,\no\\
{}^t\lambda_2^k&=&k(0,1,0,0,0,0),k=1,2,4,\no\\
{}^t\lambda_3&=&(1,1,0,0,0,0),\no\\
{}^t\lambda_4&=&(0,1,1,1,0,0),\no\\
{}^t\lambda_5&=&(0,0,0,0,1,0),\no\\
{}^t\lambda_6&=&(0,1,0,1,0,0).
\enqy

\paragraph{Level $2~E_7$ theta functions}~\\
There are 6  Weyl orbits,
\beqy
{}^t\lambda_0&=&(0,0,0,0,0,0,0),\no\\
{}^t\lambda_1&=&(1,0,0,0,0,0,0),\no\\
{}^t\lambda_2&=&(0,0,0,0,1,0,0),\no\\
{}^t\lambda_3^k&=&k(0,0,0,0,0,1,0),k=1,2,\no\\
{}^t\lambda_4&=&(0,0,0,0,0,0,1).
\enqy

\end{document}